\documentclass[11pt,a4paper,nofootinbib,superscriptaddress]{revtex4-1}
\pdfoutput=1
\usepackage[utf8]{inputenc}
\usepackage{graphicx,bm,bbm}
\usepackage{slashed,verbatim}
\usepackage{amssymb,graphicx,epstopdf}
\usepackage{slashed,subfigure}
\usepackage{caption}
\usepackage{amsmath,mathtools}
\usepackage{epstopdf,dcolumn}
\usepackage{soul}

\allowdisplaybreaks
\usepackage[normalem]{ulem}
\usepackage{cancel}
\usepackage{xcolor}
 \usepackage[colorlinks=true,linktocpage=true,linkcolor=blue,citecolor=blue]{hyperref}
\usepackage{xcolor}
\allowdisplaybreaks

\def\be {\begin{equation}}
\def\ee {\end{equation}}
\def\bea {\begin{eqnarray}}
\def\eea {\end{eqnarray}}
\def\bc {\begin{center}}
\def\ec {\end{center}}
\def\nn {\nonumber}
\def\eps {\epsilon}
\def\gm {\gamma}

\def\mn {\mu\nu}
\def\al {\alpha}
\def\({\left(}
\def\){\right)}
\def\[{\left[}
\def\]{\right]}
\def\om {\omega}
\def\sp {\shortparallel}
\newcommand \Tr{\operatorname{\text{Tr}}}

\def\sumintb{\sum\!\!\!\!\!\!\!\!\!\int\limits}

\def\slashed{\slash\!\!\!\!}

\def\sumint{\sum \!\!\!\!\!\!\!\!\int\,}

\DeclareGraphicsExtensions{.jpg,.pdf,.eps}

\begin{document}

\title{Soft contribution to the damping rate of a hard photon in a weakly magnetized hot medium}
%
 	\author{Ritesh Ghosh}
 	\email{ritesh.ghosh@saha.ac.in}
 	\author{Bithika Karmakar}
\email{bithika.karmakar@saha.ac.in}
\author{ Munshi G Mustafa}
\email{munshigolam.mustafa@saha.ac.in}
  \affiliation{
 	Theory Division, Saha Institute of Nuclear Physics, \\
 	1/AF, Bidhannagar, Kolkata 700064, India}
 	\affiliation{
 	Homi Bhabha National Institute, Anushaktinagar, \\
 	Mumbai, Maharashtra 400094, India}

\begin{abstract}
We consider weakly magnetized hot QED plasma comprising electrons and positrons. There are three distinct dispersive  (longitudinal and two transverse) modes of a photon in a thermomagnetic medium. At lowest order in the coupling constant, a photon is damped in this medium via Compton scattering and pair creation process. We evaluate the damping rate of hard photon by calculating the imaginary part of the each  transverse dispersive modes in a thermomagnetic QED medium.  We note that one of the fermions in the loop  of one-loop photon self-energy is considered as soft and the other one is hard. Considering the resummed fermion propagator in a weakly magnetized medium for the soft fermion and the Schwinger propagator for hard fermion, we calculate the soft contribution to the damping rate of hard photon.  In weak field approximation the thermal and thermomagnetic  contributions to damping rate  get separated out for each transverse dispersive mode. The total damping rate for each dispersive mode in presence of magnetic field is found to be reduced than that of the thermal one. This formalism can easily be extended to QCD plasma.
\end{abstract}

\maketitle 
\newpage
\section{Introduction}

Astrophysical plasma  is almost always immersed in magnetic field. Extreme, magnetized plasma is found in interiors of neutron star, magnetospheres of magnetars and central engines of supernovae and gamma ray bursts~\cite{Uzdensky:2014rza}. The propagation of photon through the hot magnetized plasma, \textit{viz.}, electron-positron plasma (EPP), is of great interest. Because the magnetar phenomena are found by analyzing the high-energy radiation detected at earth. Thus it is very important to have a good understanding of the propagation of photon through the EPP. Furthermore, the phenomenon of Faraday rotation {\it{i.e.,}} change of polarization of photon while propagating through a medium has been studied in Ref.~\cite{Ganguly:1999ts}
 in a field theoretical viewpoint. This has also been detected in several astrophysical objects~\cite{Giovannini:1997km}. 
 Also high-intensity laser beams are used to create ultrarelativistic EPP of temperature around 10 MeV~\cite{liang:1998}.  This EPP may play an important role in various astrophysical situations. 
 Some properties of such plasma, \textit{viz.}, the equation of state, dispersion relation of collective plasma modes of photon and electron, damping rates, mean free paths,  transport coefficients and particle production rates, are studied using QED at finite temperature~\cite{Thoma:2008my,Thoma:2008gh}.


On the other hand in noncentral heavy ion collisions, the magnetic field as high as $(15-20) m_\pi^2$ can be generated~\cite{Skokov:2009qp} at LHC energies. After a few fm/$c$ of the collision, the magnetic field strength~\footnote{The initial magnitude of this magnetic field can be very high at the initial time of the heavy-ion collisions  and then it decreases very fast, 
being inversely proportional to the square of time \cite{Bzdak:2012fr,McLerran:2013hla}. 
However for a different point of view, see Refs.~\cite{Tuchin:2012mf,Tuchin:2013bda,Tuchin:2013ie}, where the time dependence of magnetic field is shown to be adiabatic due to the high conductivity of the medium.} decreases to $(1-2) m_\pi^2$. The effect of magnetic field on the properties of the QCD matter [\textit{viz}. quark-gluon plasma(QGP)] and on the phase diagram of QCD is of great interest. Recently, several studies have found the effect of magnetic 
catalysis~\cite{Alexandre,Gusynin,Lee,Gusynin:1994re}, {\it{i.e.}}, the enhancement of phase transition temperature of QCD matter in presence of external magnetic field, whereas, some results of inverse magnetic 
catalysis~\cite{Bali:2011qj,Bornyakov:2013eya,Mueller:2015fka,Ayala:2014iba,Farias:2014eca,Ayala:2014gwa,Ayala:2016sln,Ayala:2015bgv,Farias:2016gmy} have been reported. 
Various properties of QCD matter at weak coupling in presence of magnetic field is being studied including the equation of state~\cite{Bandyopadhyay:2017cle,Karmakar:2019tdp,Rath:2017fdv},  transport properties ~\cite{Kurian:2018qwb,Kurian:2018dbn,Kurian:2017yxj}. Modification of QCD Debye mass and the two point correlation functions of quarks~\cite{Das:2017vfh} and gluons~\cite{Karmakar:2018aig,Hattori:2017xoo,Ayala:2018ina} {\it{i.e.}}, partons  have been analyzed recently. Dilepton production rate 
  from a hot magnetized QCD 
plasma~\cite{Das:2019nzv,Bandyopadhyay:2016fyd,Tuchin:2013prc,Tuchin:2013prc2,Tuchin:2013ie,Sadooghi:2016jyf,Bandyopadhyay:2017raf,Mamo:2013efa} has been calculated. The photon is also considered as a good probe of the QGP medium as photon only interacts electromagnetically. Thus, it comes out of the hot QCD system without interacting much. The damping rate of the hard photon is associated with the mean free path of photon~\cite{Thoma:1993vs} and hard photon production rate in QGP~\cite{Kapusta:1991qp}.  
    
  Damping rate of photon is related to the imaginary part of photon dispersion in the medium~\cite{Thoma:1994fd} which is again related to the scattering crosssection of the process that we find by cutting the photon self-energy diagram~\cite{Weldon:1983jn}. In lowest order coupling constant, photons are damped by Compton scattering and pair creation process. In case of low momentum transfer, the damping rate shows infrared singularity. Thus one should consider the effective resummed propagator instead of bare propagator for soft momentum of fermion. We will call this as the soft contribution to the damping rate of photon. The hard contribution refers to the case where all the fermions in loop have momentum order of or much greater than the system temperature $T$. Both soft and hard contributions to the damping rate of hard  photon in thermal medium have been calculated in Ref.~\cite{Thoma:1994fd}. The dispersion relations of photon are modified for a hot magnetized medium~\cite{Karmakar:2018aig}. So the damping rate of photon will also get modified in a thermomagnetic medium. In this article we intend to compute the soft contribution to the hard photon damping rate for a weakly magnetized hot medium in one loop approximation of photon self-energy. In a thermomagnetic medium this would be a good indicator as higher loop calculation contributing to higher order would be extremely involved.
  
  We consider hard photon of momentum $P^\mu=(p_0,\bf p)$ where $p=|\bf p|\gg $ $T$ in a relativistic hot magnetized QED medium. To find the soft contribution of the damping rate we introduce a separation scale $\Lambda$ where $eT\ll \Lambda\ll T$ \,($gT\ll \Lambda\ll T$ in case of QCD). In the soft part of the damping rate, the contribution from soft loop momentum involving a fermion is taken into account up to the separation scale $\Lambda$ . Here we assume that the magnetic field strength is weak {\it{i.e.,}} $\sqrt{eB}<eT<T$\,($\sqrt{q_fB}<gT<T$ for QCD). We use the recently obtained effective fermion  propagator~\cite{Das:2017vfh} in presence of weak magnetic field for the soft fermion and Schwinger propagator for the hard fermion in the loop. The Braaten-Pisarki-Yuan formalism~\cite{Braaten:1990wp} has been used here to calculate the imaginary part of photon self-energy. Extension to the case of damping rate of hard photon in weakly magnetized hot QCD medium is straightforward. We need to consider the loop fermions as quark and antiquark in that case.
  
  In Sec.~\ref{Formalism} we describe the set up to calculate the photon damping rate associated with imaginary part of photon self-energy. In  Sec.~\ref{self_energy_calculation} the self-energy is obtained in a weak field approximation. The imaginary parts of various components of photon  self-energy is obtained in Sec.~\ref{imaginary_part}. Results are given in Sec.~\ref{Res}. We conclude in Sec.~\ref{Conclusion}.

\section{SetUp}
\label{Formalism}
We consider plasma of electrons and positrons at temperature $T$. The $z$-axis of the lab frame is oriented along the magnetic field. 
The general structure of the gauge boson self-energy and corresponding effective propagator have been evaluated in Ref.~\cite{Karmakar:2018aig}. 
The general covariant structure of photon self-energy in a magnetized hot medium can be written as
\bea
\Pi^{\mu\nu}&=& \beta B^{\mu\nu}+\sigma R^{\mu\nu}+\delta Q^{\mu\nu}+\al N^{\mu\nu}\,\,,
\eea
where various form factors can be written as
\bea
\beta&=&B^{\mu\nu}\Pi_{\mu\nu},\nn\\
\sigma&=&R^{\mu\nu}\Pi_{\mu\nu},\nn\\
\delta&=&Q^{\mu\nu}\Pi_{\mu\nu},\nn\\
\al&=&\frac{1}{2}N^{\mu\nu}\Pi_{\mu\nu}.\label{ff}
\eea

The general covariant structure of photon propagator can be obtained~\cite{Karmakar:2018aig} as 
\bea
D_{\mu\nu}&=&\frac{\xi P_{\mu}P_{\nu}}{P^4}+\frac{(P^2-\delta)B_{\mu\nu}}{(P^2-\beta)(P^2-\delta)-\al^2}+\frac{R_{\mu\nu}}{P^2-\sigma}+\frac{(P^2-\beta)Q_{\mu\nu}}{(P^2-\beta)(P^2-\delta)-\al^2}\nn\\
&&+\frac{\al N_{\mu\nu}}{(P^2-\beta)(P^2-\delta)-\al^2}.
\label{a}
\eea
We note that the thermal medium (absence of magnetic field) has two dispersive modes of photon \textit{i.e.}, one degenerate transverse mode and one medium induced plasmon mode due to breaking of boost invariance. Now breaking of rotational invariance in the presence of a magnetic field leads to three dispersive modes of photon by lifting the degeneracy of the transverse modes. These three dispersive modes can be seen from the pole of Eq. (\ref{a}). Now, the dispersion relations can be written as 
\bea
P^2-\sigma&=&0,\\
(P^2-\delta)(P^2-\beta)-\alpha^2&=&\bigg(P^2-\frac{\beta+\delta+\sqrt{(\beta-\delta)^2+4\alpha^2}}{2}\bigg)\nn\\
&&\times \bigg(P^2-\frac{\beta+\delta-\sqrt{(\beta-\delta)^2+4\alpha^2}}{2}\bigg)=0\,.
\eea
In weak magnetic field approximation $\alpha$ does not contribute upto $\mathcal{O}[(eB)]^2$, one gets simple form of the above dispersive modes~\cite{Bandyopadhyay:2017cle}
\bea
P^2-\sigma&=&0,\nn\\
P^2-\beta&=&0,\nn\\
P^2-\delta&=&0. \, 
\eea
Damping rate is defined as the imaginary part of photon dispersion relation. The medium induced longitudinal (plasmon) mode does not contribute to the damping rate \footnote{ The longitudinal dispersive mode merges with the light cone at high photon momentum.} and
the dispersion relations for two transverse modes of a photon are given, respectively,  as
\bea
P^2-\sigma=0\,\,,\,\,P^2-\delta=0 ,
\eea
Damping rates $\gm_{\delta}(p)$ and $\gm_{\sigma}(p)$ (for no overdamping ${\it{i.e.}}$ $\gm_{i}\ll p_0\,\,\mathrm{where}\,\,\,i=\delta,\sigma$) of hard photon are given by imaginary part of the form factors as  ~\cite{Thoma:2000dc}
\bea
\gm_{\sigma}(p)&=&-\frac{1}{2p}\mathrm{Im}\,\sigma(p_0=p), \label{gamasig}\\
\gm_{\delta}(p)&=&-\frac{1}{2p}\mathrm{Im}\,\delta(p_0=p). \label{gamadel}
\eea

\noindent
The tensor structures of $R^{\mn}$ and $Q^{\mn}$~\cite{Karmakar:2018aig} are given as
\bea
R^{\mn} = 
\begin{pmatrix}
0 & 0 & 0 & 0 \\
0 & 0 & 0 & 0 \\
0 & 0 & -1 & 0 \\
0 & 0 & 0 & 0 
\end{pmatrix}
\,\,,\,\,
Q^{\mn} = 
\begin{pmatrix}
0 & 0 & 0 & 0 \\
0 & -\cos^2{\theta_p} & 0 & \sin{\theta_p}\cos{\theta_p} \\
0 & 0 & 0 & 0 \\
0 & \sin{\theta_p}\cos{\theta_p} & 0 & -\sin^2{\theta_p} \label{tensor_st}
\end{pmatrix}
.
\eea

Using Eq.\eqref{tensor_st} in Eq.(\ref{ff}) we can write the form factors $\sigma$ and $\delta$ in weak field approximation as 
\bea
\sigma&=&-\Big(\Pi^{22}_0+\Pi^{22}_2\Big),\label{sigeq}\\
\delta&=&-\cos^2{\theta_p}\Big(\Pi^{11}_0+\Pi^{11}_2\Big)-\sin^2{\theta_p}\Big(\Pi^{33}_0+\Pi^{33}_2\Big)+2 \sin{\theta_p}\cos{\theta_p}  
\Big(\Pi^{13}_0+\Pi^{13}_2\Big)\,\,. \label{deleq}
\eea
Combining Eq.\eqref{gamasig} with Eq.\eqref{sigeq}  and Eq.\eqref{gamadel} with Eq.\eqref{deleq}, the damping rates become
\bea
\gm_{\sigma}(p)&=&\frac{1}{2p}\, \Big(\mathrm{Im} \Pi^{22}_0+\mathrm{Im} \Pi^{22}_2\Big), \label{gamasig_f}\\
\gm_{\delta}(p)&=&\frac{1}{2p}\Big [ \cos^2{\theta_p}\Big(\mathrm{Im}\Pi^{11}_0+\mathrm{Im} \Pi^{11}_2\Big)
+\sin^2{\theta_p}\Big(\mathrm{Im}\ \Pi^{33}_0+ \mathrm{Im}\ \Pi^{33}_2\Big) \nn \\
&& -2 \sin{\theta_p}\cos{\theta_p}  
\Big(\mathrm{Im}\ \Pi^{13}_0+\mathrm{Im}\ \Pi^{13}_2\Big)\, \Big ]\ \label{gamadel_f}
\eea
The damping rates in Eqs.\eqref{gamasig_f} and \eqref{gamadel_f}  can now be  written as 
\bea
\gamma_\sigma(p)&=& \gamma_{\mathrm{th}}(p) + \gamma_\sigma^B(p) , \label{sig_sepa}\\
\gamma_\delta(p)&=&\gamma_{\mathrm{th}}(p) + \gamma_\delta^B(p) . \label{sig_sepa}
\eea
where $\gamma_{\mathrm{th}} $ is  the ${\cal O}[(eB)^0]$ contribution or thermal contribution is given as
\bea
\gm_{\mathrm{th}}(p)&=&\frac{1}{2p}\, \mathrm{Im} \Pi^{22}_0 =
\frac{1}{2p}\Big [ \cos^2{\theta_p} \mathrm{Im}\Pi^{11}_0
+\sin^2{\theta_p} \mathrm{Im}\ \Pi^{33}_0  -2 \sin{\theta_p}\cos{\theta_p} \mathrm{Im}\ \Pi^{13}_0\Big ].\ \label{gama_sig_f}
\eea
The thermomagnetic corrections of ${\cal O}[(eB)^2]$ are given as
\bea
\gm^B_{\sigma}(p)&=&\frac{1}{2p}\, \mathrm{Im} \Pi^{22}_2, \label{gama_sig_b}\\
\gm^B_{\delta}(p)&=&\frac{1}{2p}\Big [ \cos^2{\theta_p}\mathrm{Im} \Pi^{11}_2
+\sin^2{\theta_p} \mathrm{Im}\ \Pi^{33}_2 
 -2 \sin{\theta_p}\cos{\theta_p}  \mathrm{Im}\ \Pi^{13}_2  \Big ] . \,  \label{gama_del_b}
\eea
We need to obtain the imaginary parts of $11$, $22$, $33$ and $13$ components of the photon self-energy $\Pi^{\mn}$ which are computed in the following sections.

\section{Photon self-energy in hot magnetized medium}
\label{self_energy_calculation}

\begin{figure}[htp]
\centering
\includegraphics[width=10cm]{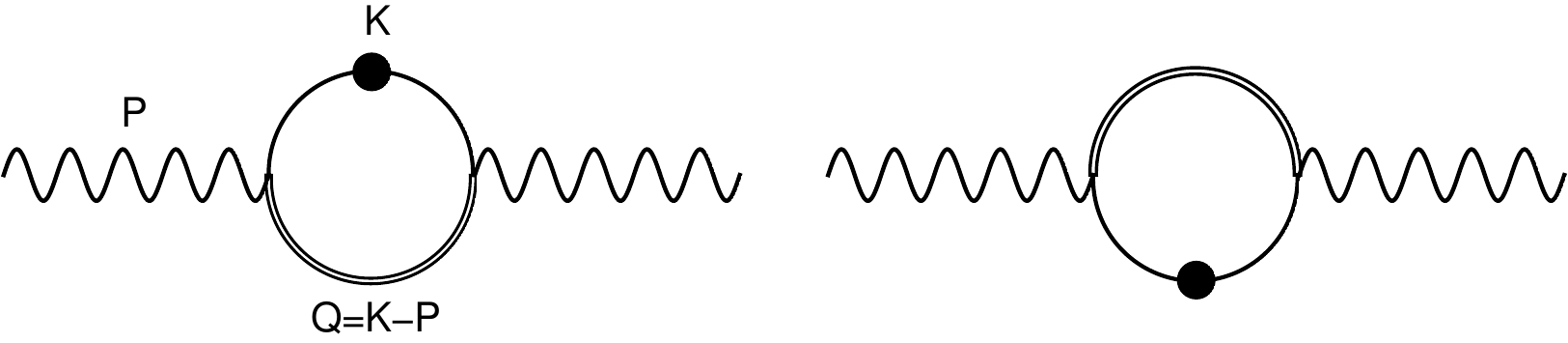}
\caption{Photon self-energy where the blob represents the effective electron propagator in magnetic field and double line 
represents bare electron propagator in magnetic field }
\label{photon_se}
\end{figure}

The photon self-energy as shown in Fig.~\ref{photon_se} can be written as 
\bea
\Pi_{\mu\nu}(P)&=&  i e^2 \int \frac{d^4 K}{(2\pi)^4}\bigg\{\Tr [\gm_{\mu}S^{*}(K)\gm_{\nu}S(Q)]+
\Tr [\gm_{\nu}S^{*}(K)\gm_{\mu}S(Q)]\bigg\}.
\label{prop}
\eea
where $S^*(K)$ is effective electron propagator and  $S(K)$ is Schwinger propagator for bare electron. As the external photon is hard, we consider one bare and one effective fermion propagator in the loop. In the following we would obtain the  propagators for fermion.

\subsection{Fermion propagator in weak field approximation}
In the weak magnetic field limit, \textit{ i.e.}, $\sqrt{eB}<m_{\mathrm{th}}\sim eT<T$ the Schwinger propagator for fermion can be written up to $\mathcal{O}[(eB)^2]$ as~\cite{Chyi:1999fc}
\bea
S(K)&=& \frac{\slashed{K}+m_f}{K^2-m_f^2}+ i\gm^1\gm^2\frac{\slashed{K_{\sp}}+m_f}{(K^2-m_f^2)^2}(eB)+  \ 2  \left[\frac{\left\{(K\cdot u)\,\slashed{u}-(K\cdot n)\,\slashed{n}\right\} 
-\slashed{K}}{(K^2-m_f^2)^3}-\frac{k_\perp^2(\slashed{K}+m_f)}{(K^2-m_f^2)^4}\right](eB)^2\nn\\
&+& \mathcal{O}\left[(e B)^3\right]\nn\\
&=& S_0+S_1+S_2+\mathcal{O}[(eB)^3].\label{schwinger_prop}
\eea
The general form of fermion self-energy in a weakly magnetized medium can be written as~\cite{Das:2017vfh}
\bea
\Sigma(K)&=& -a\,\slashed{K}-b\,\slashed{u}-b^\prime \gm_5\,\slashed{u}-c^\prime\gm_5\,\slashed{n}\,\,.
\eea
In one loop order, the form factors are given as
\bea
a(k_0,k)&=&-\frac{m_{\mathrm{th}}^2}{k^2}Q_1\bigg(\frac{k_0}{k}\bigg),\\
b(k_0,k)&=& \frac{m_{\mathrm{th}}^2}{k}\bigg[\frac{k_0}{k}Q_1\bigg(\frac{k_0}{k}\bigg)-Q_0\bigg(\frac{k_0}{k}\bigg)\bigg],\\
b^\prime(k_0,k)&=& 4 e^2 M^2(T,m_f,eB)\frac{k_3}{k^2}Q_1\bigg(\frac{k_0}{k}\bigg),\\
c^\prime(k_0,k)&=& 4 e^2 M^2(T,m_f,eB)\frac{1}{k}Q_0\bigg(\frac{k_0}{k}\bigg)\,\,,
\eea
where Legendre function of second kind are given as
\bea
Q_0(x)&=&\frac{1}{2}\ln\bigg(\frac{x+1}{x-1}\bigg), \\
Q_1(x)&=& x Q_0(x)-1,
\eea
and the thermomagnetic mass is given as
\bea
M^2(T,m_f,eB)&=&\frac{eB}{16\pi^2}\bigg[\ln 2-\frac{\pi T}{2 m_f}\bigg] ,
\label{mag_mass}
\eea
whereas thermal mass is given as
\bea
m_{\mathrm{th}}^2&=&\frac{1}{8}e^2 T^2\,\,.
\eea
The effective fermion propagator can be written~\cite{Das:2017vfh} as
\bea
S^{*}(K)&=& P_{-}\frac{\slashed{L}(K)}{L^2} P_{+}+P_{+}\frac{\slashed{R}(K)}{R^2} P_{-}\label{effective}\nn \\
&=& S^*_{L}(K)+S^*_{R}(K),
\eea
where chirality projection operators are given by
\bea
P_{\pm}=\frac{1}{2}(1\pm \gm_5),
\eea
and $L^{\mu}$ and $R^{\mu}$ are given as
\bea
L^{\mu}&=&(1+a)K^{\mu}+(b+b')u^{\mu}+c' n^{\mu},\\
R^{\mu}&=&(1+a)K^{\mu}+(b-b')u^{\mu}-c' n^{\mu}.
\eea

For simplicity of calculation we expand the effective fermion propagator in Eq.~(\ref{effective}) in powers of $eB$ and keep  terms up to $\mathcal O[(eB)^2]$ as
\bea
S^*(K)=S^*_0(K)+S^*_1(K)+S^*_{2}(K)+\mathcal O[(eB)^3],
\eea
where $S^*_0(K)$ is  ${\cal O}[(eB)^0]$ and given as
\bea
S^*_0(K)=\frac{(1+a)\,\slashed{K}+b\,\slashed{u}}{D^2}=\frac{(1+a)\,\slashed{K}+b\,\slashed{u}}{D_+ D_-}\,,
\label{htl_eff_prop}
\eea
where $D_{\pm}=(1+a)(k_0\mp k)+b$.

Equation~\eqref{htl_eff_prop} is the effective HTL fermion propagator~\cite{Bellac:2011kqa,Kapusta:2006pm} in thermal medium. The ${\cal O}[(eB)]$ is obtained as
\bea
S_1^*(K)&=&\frac{1}{D^4}\bigg[2(1+a)\slashed{K}\gm_5\bigg\{-(1+a)k_3\,c'-(1+a)k_0\, b'-bb'\bigg\}\nn\\
&+&\slashed{u}\gm_5\bigg\{\Big((1+a)^2K^2-b^2\Big)b'-2(a+1)bc'\,k_3\bigg\}\nn\\
&+&c'\slashed{n}\gm_5\bigg\{\Big(2(1+a)k_0+b\Big)b+(a+1)^2K^2\bigg\}\bigg],
\eea
whereas ${\cal O}[(eB)^2]$ is obtained as
\bea
S^*_2(K)&=&\bigg[\frac{\Big(2b'\big\{(1+a)k_0+b\big\}+2c' k_3(1+a)\Big)^2}{D^6}-\frac{b'^2-c'^2}{D^4}\bigg]\bigg\{(1+a)\,\slash\!\!\!\! K+b \,\slash\!\!\!\! u\bigg\}\nn\\
&-&\frac{\Big(2b'\big\{(1+a)k_0+b\big\}+2c' k_3(1+a)\Big)\Big(b' \,\slash\!\!\!\!u+c'\,\slash\!\!\!\! n\Big)}{D^4}\nn\\
&=&\bigg(\frac{h^2(k_0,k_{\perp},k_3)}{D^6}-\frac{h'}{D^4}\bigg)\Big\{(1+a)\,\slash\!\!\!\! K+b \,\slash\!\!\!\! u\Big\}-\frac{h(k_0,k_{\perp},k_3)}{D^4}\Big(b' \,\slash\!\!\!\!u+c'\,\slash\!\!\!\! n\Big),
\eea
where $h=2b'\big\{(1+a)k_0+b\big\}+2c' k_3(1+a)$ and $h'=b'^2-c'^2$.

\subsection{Photon self-energy in weak magnetic field}

Now the $\mathcal{O} [(eB)^0]$ contribution of $\Pi^{\mn}$ given in Eq.~(\ref{prop}) can be written as
\bea
\Pi^{\mn}_0&=&i e^2 \int \frac{d^4 K}{(2\pi)^4}\bigg\{\Tr [\gm_{\mu}S^{*}_{0}(K)\gm_{\nu}S_0(Q)]+
\Tr [\gm_{\nu}S^{*}_{0}(K)\gm_{\mu}S_0(Q)]\bigg\}\nn\\
&=&i e^2 \int \frac{d^4 K}{(2\pi)^4} \Bigg\{ \Tr\bigg[\gamma^\mu \bigg(\frac{\gamma_0-\vec \gamma \cdot \hat k}{2D_+}+\frac{\gamma_0+\vec \gamma \cdot \hat k}{2D_-}\bigg)\gamma^\nu \bigg(f_0^{(1)}\gamma^0-f_0^{(0)}\vec \gamma \cdot \vec q\bigg)\bigg]\nn\\
&+&\Tr\bigg[\gamma^\nu \bigg(\frac{\gamma_0-\vec \gamma \cdot \hat k}{2D_+}+\frac{\gamma_0+\vec \gamma \cdot \hat k}{2D_-}\bigg)\gamma^\mu \bigg(f_0^{(1)}\gamma^0-f_0^{(0)}\vec \gamma \cdot \vec q\bigg)\bigg]\Bigg\}\nn\\
&=&8ie^2\int \frac{d^4K}{(2\pi)^4}\frac{(1+a) \big\{\big(K^\mu Q^\nu +K^\nu Q^\mu\big)-g^{\mn} K\cdot Q\big\}+b \,\big\{\big(Q^\mu u^\nu+Q^\nu u^\mu\big)-g^{\mn} Q\cdot u\big\}}{D_+D_-\,Q^2},\nn\\
\label{Pimn0}
\eea
where
\bea
f_0^{(1)}&=&\frac{q_0}{Q^2}\,\,,\,\,f_0^{(0)}=\frac{1}{Q^2},\nn\\
f_1^{(1)}&=&\frac{q_0}{Q^4}\,\,,\,\,f_1^{(0)}=\frac{1}{Q^4}.
\eea

The $\mathcal{O} [(eB)]$ contribution of $\Pi^{\mn}$ is given as
\bea
\Pi^{\mn}_1&=&i e^2 \int \frac{d^4 K}{(2\pi)^4}\bigg\{
\Tr [\gm_{\mu}S^{*}_{0}(K)\gm_{\nu}S_1(Q)]+
\Tr [\gm_{\nu}S^{*}_{0}(K)\gm_{\mu}S_1(Q)]\nn\\
&+&\Tr [\gm_{\mu}S^{*}_{1}(K)\gm_{\nu}S_0(Q)]+\Tr [\gm_{\nu}S^{*}_{1}(K)\gm_{\mu}S_0(Q)]\bigg\},\label{Pimn1}
\eea
which becomes zero.

The $\mathcal{O} [(eB)^2]$ contribution of $\Pi^{\mn}$ is given as
\bea
\Pi^{\mn}_2&=&i e^2 \int \frac{d^4 K}{(2\pi)^4}\bigg\{\Tr [\gm_{\mu}S^{*}_1(K)\gm_{\nu}S_1(Q)]+\Tr [\gm_{\nu}S^{*}_1(K)\gm_{\mu}S_1(Q)]+
\Tr [\gm_{\mu}S^{*}_{0}(K)\gm_{\nu}S_2(Q)]\nn\\
&+&\Tr [\gm_{\nu}S^{*}_{0}(K)\gm_{\mu}S_2(Q)]+
\Tr [\gm_{\mu}S^{*}_2(K)\gm_{\nu}S_0(Q)]+\Tr [\gm_{\nu}S^{*}_2(K)\gm_{\mu}S_0(Q)]\bigg\}.\label{Pimn2}
\eea
We calculate the above mentioned trace as follows. The trace of the first and second terms of Eq.~\eqref{Pimn2} can be calculated as
\bea
&&\Tr [\gm_{\mu}S^{*}_1(K)\gm_{\nu}S_1(Q)]+\Tr [\gm_{\nu}S^{*}_1(K)\gm_{\mu}S_1(Q)]\nn\\
&=&\frac{8\,(eB)}{D^2\Big(Q^2-m_f^2\Big)^2}\Bigg[b'\Big\{(u^\mu n^\nu+u^\nu n^\mu)(Q\cdot u)-2u^\mu u^\nu (Q\cdot n)+g^{\mn}(Q\cdot n)\Big\}\nn\\
&+&c'\Big\{2n^\mu n^\nu (Q\cdot u)-(n^\mu u^\nu +n^\nu u^\mu)(Q\cdot n)+g^{\mn}(Q\cdot u)\Big\}\Bigg]-\frac{8\,(eB)}{D^4\Big(Q^2-m_f^2\Big)^2}\nn\\
&\times&\Bigg[h \bigg\{\big(1+a\big)\bigg(g^{\mn}\Big((K\cdot u)( Q \cdot n)-(K \cdot n)( Q\cdot u)\Big)-(K^\mu u^\nu+K^\nu u^\mu) Q\cdot n\nn\\
&+&(K^\mu n^\nu+K^\nu n^\mu)Q\cdot u\bigg)+b\bigg(g^{\mn} Q\cdot n+(u^\mu n^\nu+u^\nu n^\mu) Q\cdot u-2u^\mu u^\nu Q\cdot n\bigg)\bigg\}\Bigg].\label{Trace1}
\eea
The trace of third and fourth terms in Eq.~\eqref{Pimn2}  can be obtained as
\bea
&&\Tr [\gm_{\mu}S^{*}_{0}(K)\gm_{\nu}S_2(Q)]+\Tr [\gm_{\nu}S^{*}_{0}(K)\gm_{\mu}S_2(Q)]\nn\\
&=& \frac{8(eB)^2}{D_+(Q^2-m_f^2)^3}\bigg[q_0\Big(\hat K^\mu g^{0\nu}+\hat K^\nu g^{0 \mu}-g^{\mn}\Big)-q_3\Big(\hat K^\mu g^{3 \nu }+\hat K^\nu g^{3\mu}-g^{\mn}\hat k_3\Big)\nn\\
&-&\Big(\hat K^\mu Q^\nu + \hat K^\nu Q^\mu-g^{\mn}\hat K\cdot Q\Big)\bigg]-\frac{8(eB)^2q_{\perp}^2}{D_+(Q^2-m_f^2)^4}\bigg[\hat K^\mu Q^\nu + \hat K^\nu Q^\mu-g^{\mn}\hat K\cdot Q\bigg]\nn\\
&+&\frac{8(eB)^2}{D_-(Q^2-m_f^2)^3}\bigg[q_0\Big(\hat K'^\mu g^{0\nu}+\hat K'^\nu g^{0 \mu}-g^{\mn}\Big)-q_3\Big(\hat K'^\mu g^{3 \nu }+\hat K'^\nu g^{3\mu}+g^{\mn}\hat k_3\Big)\nn\\
&-&\Big(\hat K'^\mu Q^\nu + \hat K'^\nu Q^\mu-g^{\mn}\hat K'\cdot Q\Big)\bigg]-\frac{8(eB)^2q_{\perp}^2}{D_-(Q^2-m_f^2)^4}\bigg[\hat K'^\mu Q^\nu + \hat K'^\nu Q^\mu-g^{\mn}\hat K'\cdot Q\bigg],\label{Trace2}
\eea
where $\hat K'^\mu=(1,-\hat {\bf k})$.\\
The trace of fifth and sixth terms in Eq.~\eqref{Pimn2} are  obtained as
\bea
&&\Tr [\gm_{\mu}S^{*}_2(K)\gm_{\nu}S_0(Q)]+\Tr [\gm_{\nu}S^{*}_2(K)\gm_{\mu}S_0(Q)]\nn\\
&=& \frac{8}{(Q^2-m_f^2)}\Bigg[\Big(\frac{h^2}{D^6}-\frac{h'}{D^4}\Big)\,\Big(1+a\Big)\Big(K^\mu Q^\nu+K^\nu Q^\mu-g^{\mn}K\cdot Q\Big)+\Big(b\,\Big(\frac{h^2}{D^6}-\frac{h'}{D^4}\Big)-b'\, \frac{h}{D^4}\Big)\times\nn\\
&&\Big(g^{\mu 0} Q^\nu+g^{\nu 0} Q^\mu-g^{\mn}q_0\Big)
-c'\,\frac{h}{D^4}\Big(g^{\mu 3} Q^\nu+g^{\nu 3} Q^\mu-g^{\mn}q_3\Big)\Bigg].\label{Trace3}
\eea
The photon self-energy in weak field approximation now can be decomposed  using Eqs.\eqref{Pimn0},\eqref{Pimn1},\eqref{Pimn2} as
\bea
\Pi^{\mn}(P)= \Pi^{\mn}_0(P) + \Pi^{\mn}_2(P) , \label{decompose}
\eea
where the  first term  is  a pure thermal($\mathcal O[(eB)^0]$) contribution and  second term is  thermomagnetic correction of $\mathcal O[(eB)^2]$. 

Now the $\mathcal O[(eB)^0]$ expression of $\Pi^{11}$  , $\Pi^{22}$, $\Pi^{33}$, and $\Pi^{13}$ can be written from Eq. (\ref{Pimn0}) as
\bea
\Pi^{11}_0(p_0,p)&=&8ie^2\int \frac{d^4K}{(2\pi)^4}\frac{(1+a)(k_0q_0+2k_1q_1-\vec k\cdot \vec q)+bq_0}{\Big\{{\big((1+a) k_0+b\big)^2-(1+a)^2 k^2}\Big\}\,Q^2}\nn\\
&=&-4e^2\sumint \Bigg[\bigg(\frac{f_0^{(1)}}{D_+}+\frac{f_0^{(1)}}{D_-}\bigg)+\big(2 \hat k_{1} q_{1}-\hat k \cdot q\big) \,\, \bigg\{\frac{f_0^{(0)}}{D_+}-\frac{f_0^{(0)}}{D_-}\bigg\}\Bigg],\nn\\
\Pi^{22}_0(p_0,p)&=&8ie^2\int \frac{d^4K}{(2\pi)^4}\frac{(1+a)(k_0q_0+2k_2q_2-\vec k\cdot \vec q)+bq_0}{\Big\{{\big((1+a) k_0+b\big)^2-(1+a)^2 k^2}\Big\}\,Q^2}\nn\\
&=&-4e^2\sumint\Bigg[\bigg(\frac{f_0^{(1)}}{D_+}+\frac{f_0^{(1)}}{D_-}\bigg)+\big(2 \hat k_{2} q_{2}-\hat k \cdot q\big) \,\, \bigg\{\frac{f_0^{(0)}}{D_+}-\frac{f_0^{(0)}}{D_-}\bigg\}\Bigg],\nn\\
\Pi^{33}_0(p_0,p)&=&8ie^2\int \frac{d^4K}{(2\pi)^4}\frac{(1+a)(k_0q_0+2k_3q_3-\vec k\cdot \vec q)+bq_0}{\Big\{{\big((1+a) k_0+b\big)^2-(1+a)^2 k^2}\Big\}\,Q^2}\nn\\
&=&-4e^2\sumint \Bigg[\bigg(\frac{f_0^{(1)}}{D_+}+\frac{f_0^{(1)}}{D_-}\bigg)-\big( \hat k\cdot  q-2\hat k_3 q_3\big) \,\, \bigg\{\frac{f_0^{(0)}}{D_+}-\frac{f_0^{(0)}}{D_-}\bigg\}\Bigg],\nn\\
\Pi^{13}_0(p_0,p)&=&8ie^2 \int \frac{d^4K}{(2\pi)^4}\frac{(1+a)(k_1q_3+q_1 k_3)}{\Big\{{\big((1+a) k_0+b\big)^2-(1+a)^2 k^2}\Big\}\,Q^2}\nn\\
&=&-4e^2\sumint \Bigg[ (\hat k_1q_3+q_1 \hat k_3)\,\, \bigg\{\frac{f_0^{(0)}}{D_+}-\frac{f_0^{(0)}}{D_-}\bigg\}\Bigg].
\eea
Using Eqs.\eqref{Pimn2},\eqref{Trace1},\eqref{Trace2} and \eqref{Trace3}, one can write the $\mathcal O[(eB)^2]$ expression of $\Pi^{11}$ , 
$\Pi^{22}$, $\Pi^{33}$, and $\Pi^{13}$ as
\bea
\Pi^{11}_2&=&-e^2 \sumint \Bigg[- \frac{8 eB}{D^2(Q^2-m_f^2)^2}\bigg\{ b' q_3+c' q_0\bigg\}+\frac{8 eB}{D^4(Q^2-m_f^2)^2}\,h \bigg\{(1+a)(k_0q_3-k_3q_0)+b q_3  \bigg\}\nn\\
&+&\frac{8 (eB)^2}{D_+(Q^2-m_f^2)^3}\Big(\hat k_2 q_2-\hat k_1 q_1 \Big)-\frac{8 (eB)^2 q_{\perp}^2}{D_+(Q^2-m_f^2)^4}\Big(q_0-\hat k \cdot q +2  \hat k_1 q_1 \Big)\nn\\
&-&\frac{8 (eB)^2}{D_-(Q^2-m_f^2)^3}\Big(\hat k_2 q_2-\hat k_1 q_1 \Big)-\frac{8 (eB)^2 q_{\perp}^2}{D_-(Q^2-m_f^2)^4}\Big(q_0+\hat k \cdot q -2  \hat k_1 q_1 \Big)\nn\\
&+& \frac{8}{(Q^2-m_f^2)}\bigg\{ \Big(\frac{h^2}{D^6}-\frac{h'}{D^4}\Big)(1+a)(2k_1q_1+ K \cdot Q)+\bigg(b\Big(\frac{h^2}{D^6}-\frac{h'}{D^4}\Big)-\frac{b' h}{D^4}\bigg)q_0\nn\\
&-&\frac{c' h}{D^4} q_3 \bigg\} \Bigg],
\eea
\bea
\Pi^{22}_2&=&-e^2 \sumint \Bigg[- \frac{8 eB}{D^2(Q^2-m_f^2)^2}\bigg\{ b' q_3+c' q_0\bigg\}+\frac{8 eB}{D^4(Q^2-m_f^2)^2}\,h \bigg\{(1+a)(k_0q_3-k_3q_0)+b q_3  \bigg\}\nn\\
&+&\frac{8 (eB)^2}{D_+(Q^2-m_f^2)^3}\Big(\hat k_1 q_1-\hat k_2 q_2 \Big)-\frac{8 (eB)^2 q_{\perp}^2}{D_+(Q^2-m_f^2)^4}\Big(q_0-\hat k \cdot q +2  \hat k_2 q_2 \Big)\nn\\
&-&\frac{8 (eB)^2}{D_-(Q^2-m_f^2)^3}\Big(\hat k_1 q_1-\hat k_2 q_2 \Big)-\frac{8 (eB)^2 q_{\perp}^2}{D_-(Q^2-m_f^2)^4}\Big(q_0+\hat k \cdot q -2  \hat k_2 q_2 \Big)\nn\\
&+& \frac{8}{(Q^2-m_f^2)}\bigg\{ \Big(\frac{h^2}{D^6}-\frac{h'}{D^4}\Big)(1+a)(2k_2q_2+ K \cdot Q)+\bigg(b\Big(\frac{h^2}{D^6}-\frac{h'}{D^4}\Big)-\frac{b' h}{D^4}\bigg)q_0\nn\\
&-&\frac{c' h}{D^4} q_3 \bigg\} \Bigg],
\eea
\bea
\Pi^{33}_2&=&-e^2 \sumint \Bigg[ \frac{8 eB}{D^2(Q^2-m_f^2)^2}\bigg\{ -b' q_3+c' q_0\bigg\}+\frac{8 eB}{D^4(Q^2-m_f^2)^2}\,h \bigg\{(1+a)(k_0q_3+k_3q_0)+b q_3  \bigg\}\nn\\
&+&\frac{8 (eB)^2}{D_+(Q^2-m_f^2)^3}\Big(-q_3 \hat k_3+\hat k \cdot q \Big)-\frac{8 (eB)^2 q_{\perp}^2}{D_+(Q^2-m_f^2)^4}\Big(q_0-\hat k \cdot q +2 q_3 \hat k_3 \Big)\nn\\
&+&\frac{8 (eB)^2}{D_-(Q^2-m_f^2)^3}\Big(q_3 \hat k_3-\hat k \cdot q \Big)-\frac{8 (eB)^2 q_{\perp}^2}{D_-(Q^2-m_f^2)^4}\Big( q_0+ \hat k \cdot q-2q_3 \hat k_3\Big)\nn\\
&+& \frac{8}{(Q^2-m_f^2)}\bigg\{ \Big(\frac{h^2}{D^6}-\frac{h'}{D^4}\Big)(1+a)(2k_3q_3+ K \cdot Q)+\bigg(b\Big(\frac{h^2}{D^6}-\frac{h'}{D^4}\Big)-\frac{b' h}{D^4}\bigg)q_0\nn\\
&+&\frac{c' h}{D^4} q_3 \bigg\} \Bigg],
\eea
\bea
\Pi^{13}_2&=&-e^2 \sumint \Bigg[\frac{8 eB}{D^4(Q^2-m_f^2)^2}\,h (1+a)(k_1 q_0) -\frac{8 (eB)^2}{D_+(Q^2-m_f^2)^3}\hat k_3 q_1 -\frac{8 (eB)^2 q_{\perp}^2}{D_+(Q^2-m_f^2)^4}\nn\\
&\times&\Big(\hat k_1 q_3+\hat k_3 q_1 \Big)+\frac{8 (eB)^2}{D_-(Q^2-m_f^2)^3}\hat k_3 q_1 +\frac{8 (eB)^2 q_{\perp}^2}{D_-(Q^2-m_f^2)^4}\Big(\hat k_1 q_3+\hat k_3 q_1 \Big)+ \frac{8}{(Q^2-m_f^2)}\nn\\
&\times&\bigg\{ \Big(\frac{h^2}{D^6}-\frac{h'}{D^4}\Big)(1+a)\Big(k_1q_3+k_3q_1\Big)+\frac{c' h}{D^4} q_1 \bigg\} \Bigg].
\eea
\section{Imaginary parts of the components of the photon self-energy} 
\label{imaginary_part}
Before obtaining the imaginary parts, we discuss below the various approximations used in this calculation. 
\begin{enumerate}
\item We have considered the momentum of photon as hard $(p\gg T)$. The momentum of soft fermion $k\ll T$. Thus we can take the following approximations:
\bea
n_F(\om)\sim 1\,\,,\,\,n_F(p-\om)\sim e^{-p/T}\,\,,\,\,e^{-p/T}\sim 0.
\eea
\item
An upper cutoff $\Lambda(< T)$ of the soft fermion momentum $k$ has been introduced in the integrations. 
\item We consider $m_f=m_{\mathrm {th}}$ for electron.
\label{Im_Pimn}
\begin{figure}[htp]
	\centering
	\includegraphics[width=8cm]{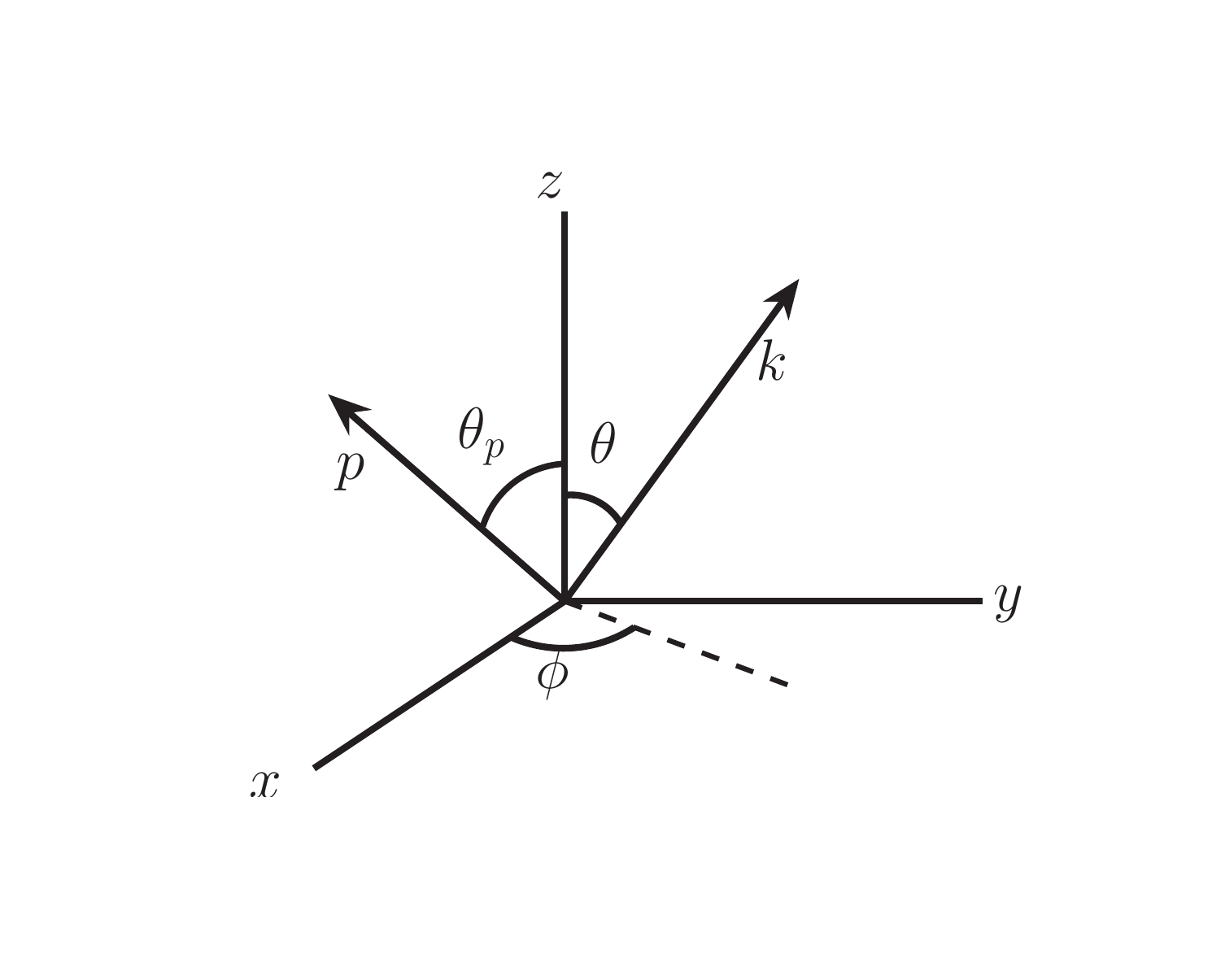}
	\caption{Choice of reference frame for computing the various  components of photon self-energy. The magnetic field is along z-direction and $\theta_p$ is the 
	angle between momentum of photon and the external magnetic field.}
	\label{frame}
\end{figure}
\item To perform the  various integrations  we choose a frame of reference as shown in Fig.~\ref{frame} in which the external momentum of the photon in $xz$ plane with 
$0<\theta_p< \pi/2$. So one can write
\be
\vec p \equiv (p \sin{\theta_p},\,0,\,p \cos{\theta_p})\,,
\ee
and then the loop momentum as
\bea
\vec k \equiv (k\sin{\theta}\cos{\phi} ,\,k\sin{\theta}\sin{\phi},\,k \cos{\theta}).
\eea
\end{enumerate}
In the following subsection we will obtain imaginary parts of various self-energy components.

\subsection{Imaginary parts of the magnetic field independent part,  $i.e.$  $\mathcal{O}[(eB)^0]$}

We evaluate the imaginary parts of $\Pi^{11}_0$  , $\Pi^{22}_0$, $\Pi^{33}_0$, and $\Pi^{13}_0$ using the Braaten-Pisarski-Yuan method~\cite{Thoma:1994fd,Braaten:1990wp}. 

\bea
\text{Im}\, \Pi^{11}_0&=&-4e^2\pi(1-e^{p/T})\int \frac{d^3k}{(2\pi)^3}\int_{-\infty}^{\infty} \int_{-\infty}^{\infty}d\omega\,d\omega' n_F(\omega) n_F(\omega')\bigg\{ \rho_0^{(1)}(\omega')\Big(\rho_{D_+}(\omega)\nn\\
&+&\rho_{D_-}(\omega)\Big)-\big(\hat k \cdot q- 2\hat k_{1} q_{1}\big) \,\,\Big(\rho_{D_+}(\omega) -\rho_{D_-}(\omega)\Big)\rho_0^{(0)}(\omega')\bigg\}\delta(\omega+\omega'-p),\label{Im_Pi110}\\
\text{Im}\, \Pi^{22}_0&=&-4e^2\pi(1-e^{p/T})\int \frac{d^3k}{(2\pi)^3}\int_{-\infty}^{\infty} \int_{-\infty}^{\infty}d\omega\,d\omega' n_F(\omega) n_F(\omega')\bigg\{ \rho_0^{(1)}(\omega')\Big(\rho_{D_+}(\omega)\nn\\
&+&\rho_{D_-}(\omega)\Big)-\big(\hat k \cdot q- 2\hat k_{2} q_{2}\big) \,\,\Big(\rho_{D_+}(\omega) -\rho_{D_-}(\omega)\Big)\rho_0^{(0)}(\omega')\bigg\}\delta(\omega+\omega'-p),\label{Im_Pi220}\\
\text{Im}\, \Pi^{33}_0&=&-4e^2\pi(1-e^{p/T})\int \frac{d^3k}{(2\pi)^3}\int_{-\infty}^{\infty} \int_{-\infty}^{\infty}d\omega\,d\omega' n_F(\omega) n_F(\omega')\bigg\{ \rho_0^{(1)}(\omega')\Big(\rho_{D_+}(\omega)\nn\\
&+&\rho_{D_-}(\omega)\Big)-\big( \hat k \cdot q-2\hat k_3 q_3\big) \,\,\Big(\rho_{D_+}(\omega) -\rho_{D_-}(\omega)\Big)\rho_0^{(0)}(\omega')\bigg\}\delta(\omega+\omega'-p),\label{Im_Pi330}\\
\text{Im}\, \Pi^{13}_0&=&-4e^2\pi(1-e^{p/T})\int \frac{d^3k}{(2\pi)^3}\int_{-\infty}^{\infty} \int_{-\infty}^{\infty}d\omega\,d\omega' n_F(\omega) n_F(\omega')\bigg\{(\hat k_1 q_3+ q_1 \hat k_3)\nn\\
&\times&\Big(\rho_{D_+}(\omega) -\rho_{D_-}(\omega)\Big)\rho_0^{(0)}(\omega')\bigg\}\delta(\omega+\omega'-p)\label{Im_Pi130},
\eea
where $\rho_0^{(0)},\rho_0^{(1)},\rho_1^{(0)},\rho_1^{(1)},\rho_{D_+}$ and $\rho_{D_-}$ are spectral representation of $f_0^{(0)},f_0^{(1)},f_1^{(0)},f_1^{(1)},1/D_+$ and $1/D_-$ respectively. These  spectral functions are obtained in  Appendix \ref{app_a}. We know that  both $\rho_{D_+}$ and $\rho_{D_-}$  have pole containing the mass shell $\delta$-function 
+ Landau cut  part in space like region whereas $\rho_0^{(0)},\rho_0^{(1)},\rho_1^{(0)},\rho_1^{(1)}$ have only pole containing the mass shell $\delta$ function. Since imaginary parts of various components of the self-energy contain the product of two spectral functions, it would then have the \textit{pole-pole} and the \textit{pole-cut} contributions.

The pole-pole parts of $\text{Im} \Pi_{11}$, $\text{Im} \Pi_{22}$, $\text{Im} \Pi_{33}$ and $\text{Im} \Pi_{13}$ contain $\delta(p-\omega_\pm -q)$ where $\omega_\pm$ is the energy of the fermion quasiparticle, $\vec k$ and $\vec q=\vec k-\vec p$ are the momenta of soft and hard fermion, respectively. Hence $\omega_\pm > k$. The $\delta$-function yields
\bea
&&p-\omega_\pm -q=0\nn\\
\cos{\phi}&\approx& \frac{\omega_\pm/k -\cos{\theta} \cos{\theta_p}}{\sin{\theta} \sin{\theta_p}} \nn. 
\eea
The value of $\frac{\omega_\pm/k  -\cos{\theta} \cos{\theta_p}}{\sin{\theta} \sin{\theta_p}}$ excludes the range $[-1,1]$ for all values of the parameters $\theta$ and $\theta_p$ . This restriction is valid for both thermal and the magnetic case. Thus pole-pole parts do not contribute in this calculation~\cite{Thoma:1994fd, Kapusta:1991qp}. In ${\cal O}[(eB)^0]$ the contribution comes  only from the \textit{pole-cut} part.

\subsubsection{Pole-cut part of $\mathcal{O}[(eB)^0]$}
Now we would find the \textit{pole-cut} part  of the above self-energy components in Eqs.\eqref{Im_Pi110}, \eqref{Im_Pi220}, \eqref{Im_Pi330} and \eqref{Im_Pi130} as

\bea
&&\text{Im}\,\Pi^{11}_0\bigg|_{pole-cut}\nn\\
&=&2e^2\pi(1-e^{p/T})\int \frac{d^3k}{(2\pi)^3}\int_{-\infty}^{\infty} \int_{-\infty}^{\infty}d\omega\,d\omega'
n_F(\om)n_F(\om')\bigg\{  \delta(\om'-q) \Theta(k^2-\om^2)\nn\\
&\times&\Big(\beta_+(\om)+\beta_-(\om)  \Big)-\frac{1}{q}(\hat k \cdot q- 2\hat k_{1} q_{1}) \delta(\om'-q) \Theta(k^2-\om^2)\Big(\beta_+(\om)-\beta_-(\om)  \Big) \bigg\}\nn\\
&\times&\delta(p-\om-\om'),\nn\\
&=&-e^2\pi \int_{0}^{\Lambda} \frac{k^2 dk}{2\pi^2} \int_{0}^{\pi}\frac{1}{2} \sin{\theta}\, d\theta \int_{0}^{2\pi}\frac{d\phi}{2\pi} \int_{-k}^{k}d\omega\bigg\{ \Big(\beta_+(\om)+\beta_-(\om)  \Big)-\frac{1}{q}(\hat k \cdot q- 2\hat k_{1} q_{1})\nn\\
&\times&\Big(\beta_+(\om)-\beta_-(\om)  \Big) \bigg\}\delta(p-\om-q),
\label{Pi_11}
\eea
\bea
&&\text{Im}\,\Pi^{22}_0\bigg|_{pole-cut}\nn\\
&=&2e^2\pi(1-e^{p/T})\int \frac{d^3k}{(2\pi)^3}\int_{-\infty}^{\infty} \int_{-\infty}^{\infty}d\omega\,d\omega'
n_F(\om)n_F(\om')\bigg\{  \delta(\om'-q) \Theta(k^2-\om^2)\nn\\
&\times&\Big(\beta_+(\om)+\beta_-(\om)  \Big)-\frac{1}{q}(\hat k \cdot q- 2\hat k_{2} q_{2}) \delta(\om'-q) \Theta(k^2-\om^2)\Big(\beta_+(\om)-\beta_-(\om)  \Big) \bigg\}\nn\\
&\times&\delta(p-\om-\om')\nn\\
&=&-e^2\pi \int_{0}^{\Lambda} \frac{k^2 dk}{2\pi^2} \int_{0}^{\pi}\frac{1}{2} \sin{\theta}\, d\theta \int_{0}^{2\pi}\frac{d\phi}{2\pi} \int_{-k}^{k}d\omega \bigg\{ \Big(\beta_+(\om)+\beta_-(\om)  \Big)-\frac{1}{q}(\hat k \cdot q- 2\hat k_{2} q_{2})\nn\\
&\times&\Big(\beta_+(\om)-\beta_-(\om)  \Big) \bigg\}\delta(p-\om-q),
\label{Pi_22}
\eea
\bea
&&\text{Im}\,\Pi^{33}_0\bigg|_{pole-cut}\nn\\
&=&2e^2\pi(1-e^{p/T})\int \frac{d^3k}{(2\pi)^3}\int_{-\infty}^{\infty} \int_{-\infty}^{\infty}d\omega\,d\omega'
n_F(\om)n_F(\om')\bigg\{  \delta(\om'-q) \Theta(k^2-\om^2)\nn\\
&\times&\Big(\beta_+(\om)+\beta_-(\om)  \Big)-\frac{1}{q}(\hat k_{\perp}q_{\perp}-\hat k_3 q_3) \delta(\om'-q) \Theta(k^2-\om^2)\Big(\beta_+(\om)-\beta_-(\om)  \Big) \bigg\}\nn\\
&\times&\delta(p-\om-\om')\nn\\
&=&-e^2\pi \int_{0}^{\Lambda} \frac{k^2 dk}{2\pi^2} \int_{0}^{\pi}\frac{1}{2} \sin{\theta}\, d\theta \int_{0}^{2\pi}\frac{d\phi}{2\pi} \int_{-k}^{k}d\omega \bigg\{ \Big(\beta_+(\om)+\beta_-(\om)  \Big)-\frac{1}{q}(\hat k_{\perp}q_{\perp}-\hat k_3 q_3)\nn\\
&\times&\Big(\beta_+(\om)-\beta_-(\om)  \Big) \bigg\}\delta(p-\om-q),
\label{Pi_33}
\eea
\bea
&&\text{Im}\,\Pi^{13}_0\bigg|_{pole-cut}\nn\\
&=&2e^2\pi(1-e^{p/T})\int \frac{d^3k}{(2\pi)^3}\int_{-\infty}^{\infty} \int_{-\infty}^{\infty}d\omega\,d\omega'
n_F(\om)n_F(\om')\bigg\{\frac{\hat k_1q_3 +q_1 \hat k_3}{q}\,\delta(\om'-q)\nn\\
&\times&\Big(\beta_+(\om)-\beta_-(\om)  \Big)\Theta(k^2-\om^2)   \bigg\}\delta(p-\om-\om')\nn\\
&=&-e^2\pi \int_{0}^{\Lambda} \frac{k^2 dk}{2\pi^2} \int_{0}^{\pi}\frac{1}{2} \sin{\theta}\, d\theta \int_{0}^{2\pi}\frac{d\phi}{2\pi} \int_{-k}^{k}d\omega \,\frac{1}{q}\Big( \hat k_1q_3 +q_1 \hat k_3\Big) \Big(\beta_+(\om)-\beta_-(\om)  \Big)  \nn\\
&\times&\delta(p-\om-q).\nn\\
 \label{Pi_13}
\eea
\noindent
Here we note that the terms with $\delta(\om'+q)\, \delta(p-\om-\om')\, \Theta(k^2-\om^2)$ will not contribute because $k^2-(p+q)^2$ can not be greater than zero. 
So we have excluded  those terms.

\subsection{Imaginary part of magnetic field dependent part of $\mathcal{O}[(eB)^2]$}
\noindent
Similar to $\mathcal O[(eB)^0]$ case, the 
imaginary part of $\Pi^{11}_2$, $\Pi^{22}_2$, $\Pi^{33}_2$ and $\Pi^{13}_2$ can be written as
\bea
\text{Im}\, \Pi_2^{11}
&=&-8e^2\pi(1-e^{p/T})\int \frac{d^3k}{(2\pi)^3}\int_{-\infty}^{\infty} \int_{-\infty}^{\infty}d\omega\,d\omega' n_F(\omega)\,n_F(\omega')\nn\\ 
&\times&\bigg[eB\Big\{q_3\,\rho_1^{(0)}\Big(\rho_9^{(1)}+\rho_{10}-\rho_7\Big)-\rho_1^{(1)}\Big(\rho_8+{k_3}\,\rho_9^{(0)}\Big)\Big\}+(eB)^2\Big(\hat k_1 q_1-\hat k_2 q_2\Big)\rho_2^{(0)}\nn\\
&\times&\Big(\rho_{D_-}-\rho_{D_+}\Big)-(eB)^2q_\perp^2\Big\{\rho_3^{(1)}\Big(\rho_{D_+}+\rho_{D_-}\Big)+(2\hat k_1 q_1-\hat k\cdot q)\rho_3^{(0)}\Big(\rho_{D_+}-\rho_{D_-}\Big)\Big\}\nn\\
&+&\rho_0^{(1)}\Big(\rho_{15}^{(1)}-\rho_{14}^{(1)}+\rho_{16}-\rho_{13}-\rho_{11}\Big)+(2k_1\,q_1-\vec k \cdot \vec q)\rho_0^{(0)}\Big(\rho_{15}^{(0)}-\rho_{14}^{(0)}\Big)-q_3\,\rho_0^{(0)}\rho_{12}\bigg]\nn\\
&\times&\delta(p-\om-\om'),\nn\\
\label{ImPiB11}
\eea
\bea
\text{Im}\, \Pi_2^{22}
&=&-8e^2\pi(1-e^{p/T})\int \frac{d^3k}{(2\pi)^3}\int_{-\infty}^{\infty} \int_{-\infty}^{\infty}d\omega\,d\omega' n_F(\omega)\,n_F(\omega')\nn\\ 
&\times&\bigg[eB\Big\{q_3\,\rho_1^{(0)}\Big(\rho_9^{(1)}+\rho_{10}-\rho_7\Big)-\rho_1^{(1)}\Big(\rho_8+{k_3}\,\rho_9^{(0)}\Big)\Big\}+(eB)^2\Big(\hat k_2 \,q_2-\hat k_1 q_1\Big)\rho_2^{(0)}\nn\\
&\times&\Big(\rho_{D_-}-\rho_{D_+}\Big)-(eB)^2q_\perp^2\Big\{\rho_3^{(1)}\Big(\rho_{D_+}+\rho_{D_-}\Big)+(2\hat k_2\, q_2-\hat k\cdot q)\rho_3^{(0)}\Big(\rho_{D_+}-\rho_{D_-}\Big)\Big\}\nn\\
&+&\rho_0^{(1)}\Big(\rho_{15}^{(1)}-\rho_{14}^{(1)}+\rho_{16}-\rho_{13}-\rho_{11}\Big)+(2k_2\,q_2-\vec k \cdot \vec q)\rho_0^{(0)}\Big(\rho_{15}^{(0)}-\rho_{14}^{(0)}\Big)-q_3\,\rho_0^{(0)}\rho_{12}\bigg]\nn\\
&\times&\delta(p-\om-\om'),\nn\\
\label{ImPiB22}
\eea
\bea
\text{Im}\, \Pi_2^{33}
&=&-8e^2\pi(1-e^{p/T})\int \frac{d^3k}{(2\pi)^3}\int_{-\infty}^{\infty} \int_{-\infty}^{\infty}d\omega\,d\omega' n_F(\omega)\,n_F(\omega')\nn\\ 
&\times&\bigg[eB\Big\{q_3\,\rho_1^{(0)}\Big(\rho_9^{(1)}+\rho_{10}-\rho_7\Big)+\rho_1^{(1)}\Big(\rho_8+{k_3}\,\rho_9^{(0)}\Big)\Big\}+(eB)^2\Big(q_3\hat k_3-\hat k \cdot q\Big)\rho_2^{(0)}\nn\\
&\times&\Big(\rho_{D_-}-\rho_{D_+}\Big)-(eB)^2q_\perp^2\Big\{\rho_3^{(1)}\Big(\rho_{D_+}+\rho_{D_-}\Big)+(2q_3\,\hat k_3-\hat k\cdot q)\rho_3^{(0)}\Big(\rho_{D_+}-\rho_{D_-}\Big)\Big\}\nn\\
&+&\rho_0^{(1)}\Big(\rho_{15}^{(1)}-\rho_{14}^{(1)}+\rho_{16}-\rho_{13}-\rho_{11}\Big)+(2k_3\,q_3-\vec k \cdot \vec q)\rho_0^{(0)}\Big(\rho_{15}^{(0)}-\rho_{14}^{(0)}\Big)+q_3\,\rho_0^{(0)}\rho_{12}\bigg]\nn\\
&\times&\delta(p-\om-\om'),\nn\\
\label{ImPiB33}
\eea
\bea
\text{Im}\, \Pi_2^{13}
&=&-8e^2\pi(1-e^{p/T})\int \frac{d^3k}{(2\pi)^3}\int_{-\infty}^{\infty} \int_{-\infty}^{\infty}d\omega\,d\omega' n_F(\omega)\,n_F(\omega')\nn\\ 
&\times&\bigg[eB\Big\{\rho_1^{(1)}{k_3}\,\rho_9^{(0)}\Big\}+(eB)^2 \,\hat k_3 \,q_1 \rho_2^{(0)}\Big\{\rho_{D_-}-\rho_{D_+}\Big\}-(eB)^2q_\perp^2\Big\{(\hat k_1 q_3+\hat k_3 q_1)\rho_3^{(0)}\nn\\
&\times&\Big(\rho_{D_+}-\rho_{D_-}\Big)\Big\}+(k_1 q_3+ k_3 q_1)\rho_0^{(0)}\Big(\rho_{15}^{(0)}-\rho_{14}^{(0)}\Big)+q_1\,\rho_0^{(0)}\rho_{12}\bigg]\delta(p-\om-\om').
\label{ImPiB13}
\eea
Various spectral functions are obtained in Appendix~\ref{app_a}. As discussed before we also note that the
imaginary part of various components of the self-energy contain the \textit{pole-pole} and the \textit{pole-cut} contributions. As explained earlier the phase space does not allow the \textit{pole-pole} part to contribute in this order.  In ${\cal O}[(eB)^2]$ the contribution comes  only from  the \textit{pole-cut} part.

\subsubsection{Pole-cut part of $\mathcal{O}[(eB)^2]$}
Now the expressions of pole-cut parts of Eqs.~\eqref{ImPiB11}, \eqref{ImPiB22}, \eqref{ImPiB33} and \eqref{ImPiB13} after using the approximations, are given below:
\bea
&&\text{Im}\,\Pi^{11}_2\Big\vert_{pole-cut}\nn\\
&=&4e^2\pi \int_0^\Lambda \frac{k^2 dk}{2\pi^2}\int \frac{1}{2} \sin{\theta} \, d\theta \int \frac{d\phi}{2\pi}\int_{-k}^{k} d\omega \int_{-\infty}^{\infty}d\omega' \bigg[ \delta'''(\om'-\om_q)\bigg\{ -\frac{(eB)^2 q_{\perp}^2}{96\om_q^3}\nn\\
&\times&\Big(\beta_++\beta_-\Big)-\frac{(eB)^2 q_{\perp}^2 (2\hat k_1 q_1-\hat k \cdot q)}{96\om_q^4}\Big(\beta_+-\beta_-\Big)\bigg\} - \delta''(\om'-\om_q)\bigg\{\frac{3(eB)^2}{64\om_q^3}\nn\\
&\times&\Big(\frac{q_{\perp}^2 \,(\hat k_1 q_1-\hat k \cdot q)}{\om_q^2}-(\hat k_1q_1-\hat k_2 q_2)\Big)\times \Big(\beta_+-\beta_- \Big)+\frac{3(eB)^2q_{\perp}^2}{128\om_q^4}  \Big(\beta_+ +\beta_- \Big) \bigg\}\nn\\
&+&\delta'(\om'-\om_q)\bigg\{\frac{eB\,q_3}{4\om_q^2} \Big(\beta_9^{(1)}+\beta_{10}-\beta_{7}\Big)-\frac{eB}{4\om_q}\Big(\beta_8 +k_3 \,\beta_9^{(0)}  \Big)-\frac{(eB)^2}{16\om_q^4}\Big(\hat k_2 q_2-\hat k_1 q_1\nn\\
&+&\frac{5 q_{\perp}^2 (\hat k_1 q_1-\hat k \cdot q)}{2\om_q^2} \Big)\Big(\beta_+ -\beta_-  \Big) -\frac{(eB)^2 q_{\perp}^2}{32 \om_q^5} \Big(\beta_+ +\beta_-  \Big)\bigg\}+\delta(\om'-\om_q)\bigg\{\frac{eB\,q_3}{4\om_q^3}\nn\\
&\times&\Big( \beta_9^{(1)}+\beta_{10}-\beta_7 \Big)-\frac{(eB)^2}{16\om_q^5}\Big(3\hat k_2 q_2-3\hat k_1 q_1+\frac{5q_{\perp}^2 \,(\hat k_1 q_1-\hat k \cdot q)}{2\om_q^2} \Big)\Big(\beta_+-\beta_-  \Big)\nn\\
&-&\frac{1}{2}\bigg( \beta_{15}^{(1)}+\beta_{16}-\Big(\beta_{14}^{(1)}+\beta_{11}+\beta_{13} \Big)\bigg)-\frac{(2 k_1 q_1-\vec k \cdot \vec q\,\,)}{2\om_q}\bigg( \beta_{15}^{(0)}-\beta_{14}^{(0)} \bigg) +\frac{q_3}{2\om_q}\,\beta_{12}  \bigg\}\bigg]\nn\\
&\times&\delta(p-\om-\om'),\label{pceB11}
\eea
\bea
&&\text{Im}\,\Pi^{22}_2\Big\vert_{pole-cut}\nn\\
&=&4e^2\pi \int_0^\Lambda \frac{k^2 dk}{2\pi^2}\int \frac{1}{2} \sin{\theta} \, d\theta \int \frac{d\phi}{2\pi}\int_{-k}^{k} d\omega \int_{-\infty}^{\infty}d\omega' \bigg[ \delta'''(\om'-\om_q)\bigg\{ -\frac{(eB)^2 q_{\perp}^2}{96\om_q^3}\nn\\
&\times&\Big(\beta_+ +\beta_-\Big)-\frac{(eB)^2 q_{\perp}^2 (2\hat k_2 q_2-\hat k \cdot q)}{96\om_q^4}\Big(\beta_+-\beta_-\Big)\bigg\} - \delta''(\om'-\om_q)\bigg\{\frac{3(eB)^2}{64\om_q^3}\nn\\
&\times&\Big(\frac{q_{\perp}^2 \,(\hat k_2 q_2-\hat k \cdot q)}{\om_q^2}-(\hat k_2q_2-\hat k_1 q_1)\Big)\times \Big(\beta_+-\beta_- \Big)+\frac{3(eB)^2q_{\perp}^2}{128\om_q^4}  \Big(\beta_+ +\beta_- \Big) \bigg\}\nn\\
&+&\delta'(\om'-\om_q)\bigg\{\frac{eB\,q_3}{4\om_q^2} \Big(\beta_9^{(1)}+\beta_{10}-\beta_7\Big)-\frac{eB}{4\om_q}\Big(\beta_8 +k_3\, \beta_9^{(0)}  \Big)-\frac{(eB)^2}{16\om_q^4}\Big(\hat k_1 q_1-\hat k_2 q_2\nn\\
&+&\frac{5 q_{\perp}^2 (\hat k_2 q_2-\hat k \cdot q)}{2\om_q^2} \Big)\Big(\beta_+-\beta_-  \Big) -\frac{(eB)^2 q_{\perp}^2}{32 \om_q^5} \Big(\beta_+ +\beta_-  \Big)\bigg\}+\delta(\om'-\om_q)\bigg\{\frac{eB\,q_3}{4\om_q^3}\nn\\
&\times&\Big( \beta_9^{(1)}+\beta_{10}-\beta_7 \Big)-\frac{(eB)^2}{16\om_q^5}\Big(3\hat k_1 q_1-3\hat k_2 q_2+\frac{5q_{\perp}^2 \,(\hat k_2 q_2-\hat k \cdot q)}{2\om_q^2} \Big)\Big(\beta_+-\beta_-  \Big)\nn\\
&-&\frac{1}{2}\bigg( \beta_{15}^{(1)}+\beta_{16}-\Big(\beta_{14}^{(1)}+\beta_{11}+\beta_{13} \Big)\bigg)
-\frac{(2 k_2 q_2-\vec k \cdot \vec q\,\,)}{2\om_q}\bigg(  \beta_{15}^{(0)}-\beta_{14}^{(0)}  \bigg) +\frac{q_3}{2\om_q}\,\beta_{12}  \bigg\}\bigg] \nn\\
&\times&\delta(p-\om-\om'),\label{pceB22}
\eea
\bea
&&\text{Im}\,\Pi^{33}_2\Big\vert_{pole-cut}\nn\\
&=&4e^2\pi \int_0^\Lambda \frac{k^2 dk}{2\pi^2}\int \frac{1}{2} \sin{\theta} \, d\theta \int \frac{d\phi}{2\pi}\int_{-k}^{k} d\omega \int_{-\infty}^{\infty}d\omega' \bigg[ \delta'''(\om'-\om_q)\bigg\{ -\frac{(eB)^2 q_{\perp}^2}{96\om_q^3}\nn\\
&\times&\Big(\beta_+ + \beta_-\Big)-\frac{(eB)^2 q_{\perp}^2 (\hat k_3 q_3-\hat k_\perp q_\perp)}{96\om_q^4}\Big(\beta_+-\beta_-\Big)\bigg\} - \delta''(\om'-\om_q)\bigg\{\frac{3(eB)^2}{64\om_q^3}\nn\\
&\times&\Big(\frac{q_{\perp}^2 \,(\hat k_3 q_3-\hat k_\perp q_\perp)}{\om_q^2}+\hat k_{\perp}q_{\perp}\Big)\times \Big(\beta_+ -\beta_-\Big)+\frac{3(eB)^2q_{\perp}^2}{128\om_q^4}  \Big(\beta_+ +\beta_- \Big) \bigg\}\nn\\
&+&\delta'(\om'-\om_q)\bigg\{\frac{eB\,q_3}{4\om_q^2} \Big((\beta_9^{(1)}+\beta_{10}-\beta_7 \Big)+\frac{eB}{4\om_q}\Big(\beta_8 +k_3\, \rho_9^{(0)}  \Big)-\frac{(eB)^2}{16\om_q^4}\Big(\hat k_{\perp} q_{\perp}\nn\\
&+&\frac{5 q_{\perp}^2 (\hat k_3 q_3-\hat k_\perp q_\perp)}{2\om_q^2} \Big)\Big(\beta_+-\beta_-  \Big) -\frac{(eB)^2 q_{\perp}^2}{32 \om_q^5} \Big(\beta_+ +\beta_- \Big)\bigg\}+\delta(\om'-\om_q)\nn\\
&\times&\bigg\{\frac{eB\,q_3}{4\om_q^3}\Big( \beta_9^{(1)}+\beta_{10}-\beta_7 \Big)-\frac{(eB)^2}{16\om_q^5}\Big(3\hat k_{\perp} q_{\perp}+\frac{5q_{\perp}^2 \,(\hat k_3 q_3-\hat k _\perp q_\perp)}{2\om_q^2} \Big)\nn\\
&\times&\Big(\beta_+ -\beta_-  \Big)-\frac{1}{2}\bigg( \beta_{15}^{(1)}+\beta_{16}-\Big(\beta_{14}^{(1)}+\beta_{13}+\beta_{11} \Big)\bigg)-\frac{(2 k_3 q_3-\vec k \cdot \vec q\,\,)}{2\om_q}\nn\\
&\times&\bigg( \beta_{15}^{(0)}-\beta_{14}^{(0)} \bigg) -\frac{q_3}{2\om_q} \,\beta_{12}  \bigg\}\bigg]\delta(p-\om-\om'),\label{pceB33}
\eea
\bea
&&\text{Im}\,\Pi^{13}_2\Big\vert_{pole-cut}\nn\\
&=&4e^2\pi \int_0^\Lambda \frac{k^2 dk}{2\pi^2}\int \frac{1}{2} \sin{\theta} \, d\theta \int \frac{d\phi}{2\pi}\int_{-k}^{k} d\omega \int_{-\infty}^{\infty}d\omega' \bigg[ \delta'''(\om'-\om_q)\nn\\
&\times&\bigg\{-\frac{(eB)^2 q_{\perp}^2 (\hat k_1 q_3+\hat k_3 q_1)}{96\om_q^4}\Big(\beta_+ -\beta_-\Big)\bigg\} - \delta''(\om'-\om_q)\bigg\{\frac{3(eB)^2}{64\om_q^3}\Big(\frac{q_{\perp}^2 \,(\hat k_1 q_3+\hat k_3 q_1)}{\om_q^2}\nn\\
&-&\hat k_{3}q_{1}\Big)\Big(\beta_+-\beta_- \Big) \bigg\}+\delta'(\om'-\om_q)\bigg\{\frac{eB}{4\om_q}\,k_1 \,\beta_9^{(0)} -\frac{(eB)^2}{16\om_q^4}\Big(-\hat k_{3} q_{1}+\frac{5 q_{\perp}^2 (\hat k_1 q_3+\hat k_3 q_1)}{2\om_q^2} \Big)\nn\\
&\times&\Big(\beta_+ - \beta_-  \Big) \bigg\}+\delta(\om'-\om_q)\bigg\{-\frac{(eB)^2}{16\om_q^5}\Big(-3\hat k_{3} q_{1}+\frac{5q_{\perp}^2 \,(\hat k_1 q_3+\hat k_3 q_1)}{2\om_q^2} \Big)\Big(\beta_+ -\beta_-  \Big)\nn\\
&-&\frac{( k_1 q_3+ k_3 q_1)}{2\om_q}\bigg( \beta_{15}^{(0)}-\beta_{14}^{(0)} \bigg)-\frac{q_1}{2\om_q} \beta_{12}  \bigg\}\bigg]\delta(p-\om-\om').\label{pceB13}
\eea
\section{Results}
\label{Res}
We perform the integrations in Eq.~\eqref{Pi_11},\eqref{Pi_22},\eqref{Pi_33},\eqref{Pi_13},\eqref{pceB11},\eqref{pceB22},\eqref{pceB33} and \eqref{pceB13} numerically. In this calculation we have taken $m_\pi=0.14 \text{ GeV}$. The results are shown for $\Lambda=0.25$ GeV which satisfies $e T \ll \Lambda \ll T$.
\begin{figure}[htp]
	\centering
	\includegraphics[width=8cm]{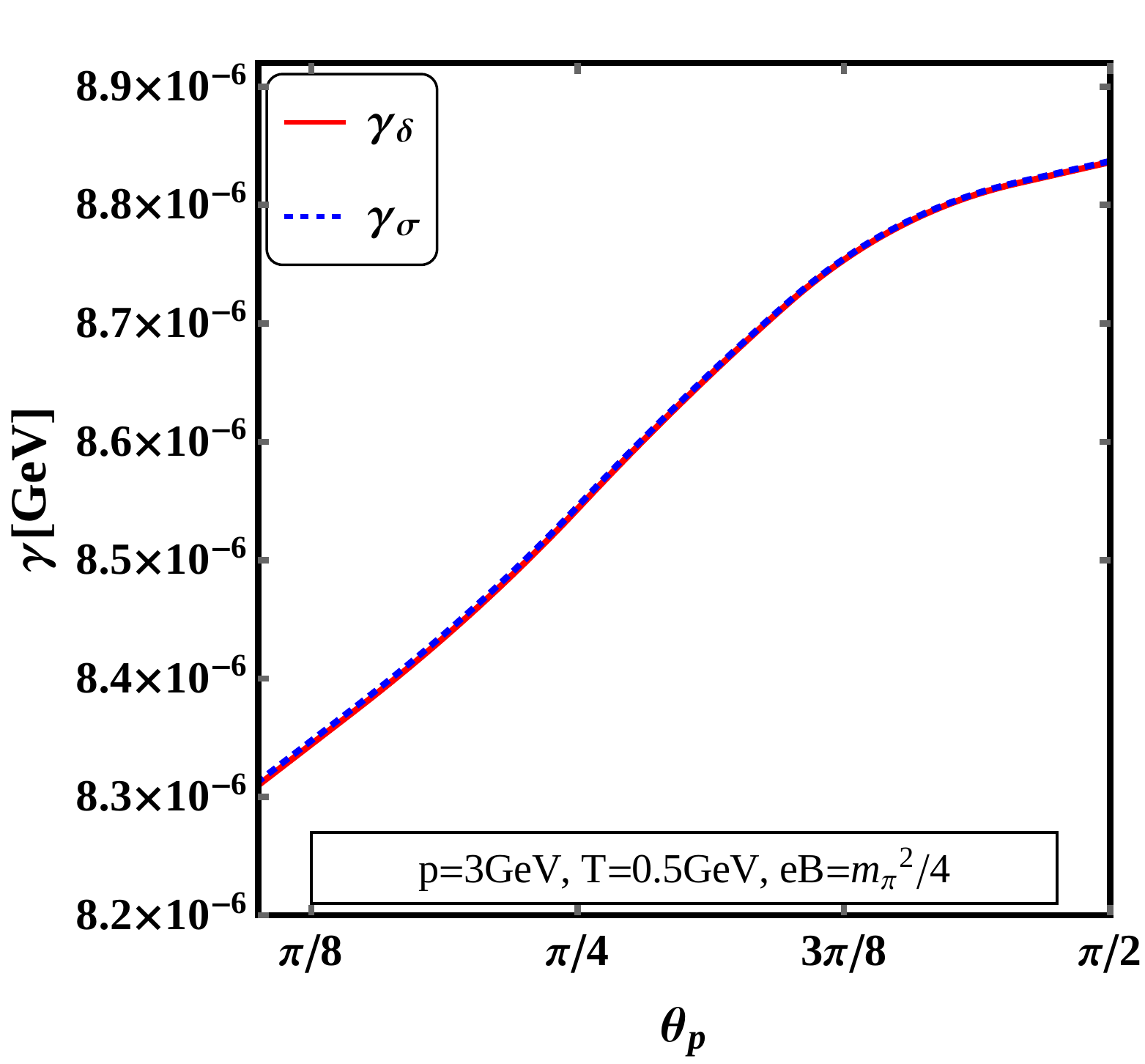}
	\caption{Plot of damping rate of photon with the propagation angle $\theta_p$ for $p=3$ GeV, $T=0.5$ GeV and $eB=m_\pi^2/4$.}
	\label{drvsang}
\end{figure}

The damping rate of photon in presence of magnetic field depends on the angle, $\theta_p$, between the momentum of photon and the magnetic field. Figure~\ref{drvsang} shows the variation of the damping rate of a hard photon with the propagation angle.  It increases with the increasing propagation angle. One can see that the two transverse modes of a hard photon are damped in a similar fashion. Since the magnetic field strength is very weak, this difference appears to be very small. We note that the magnetic correction is  $\sim \mathcal O[(eB)^2]$ and switching the magnetic field from $z$ to $-z$ direction would not affect the result. These two orientations of the magnetic field correspond to the propagation angle of photon $\theta_p$ and $\pi-\theta_p$. These two situations are identical which correspond to the same damping rates of photon at $\theta_p$ and $\pi-\theta_p$.  
\begin{figure}[htp]
	\centering
	\includegraphics[width=8cm]{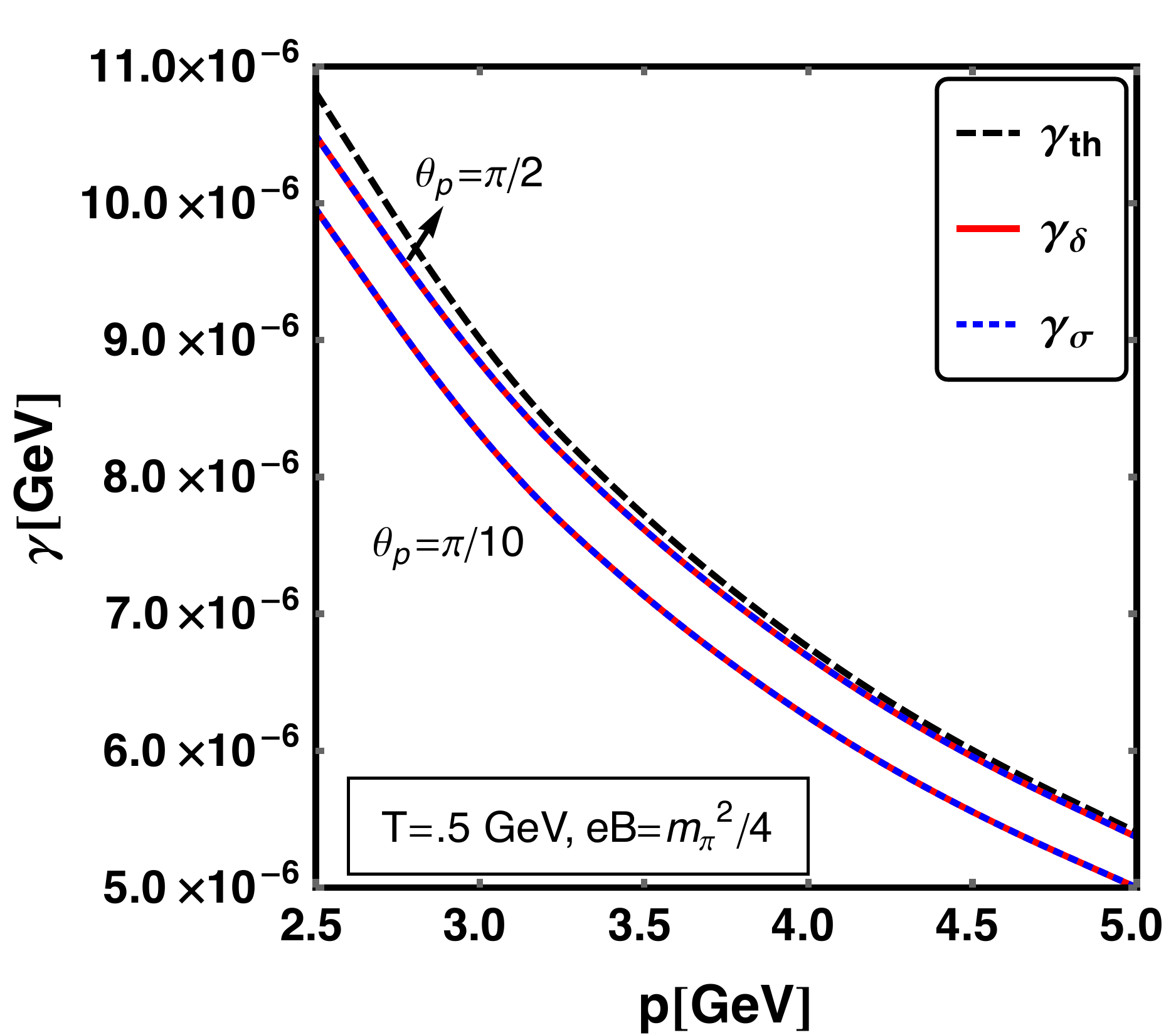}
	\caption{Plot of damping rate of photon with the energy for $T=0.5$ GeV and $eB=m_\pi^2/4$ at propagation angles $\theta_p=\pi/10$ and $\pi/2$.}
	\label{drvsp}
\end{figure}

In Fig.~\ref{drvsp} we display the damping rate as a function of photon momentum for two propagation angles $\pi/10$ and $\pi/2$. The soft contribution of the damping rate in a thermal medium agrees well with that obtained in Ref.~\cite{Thoma:1994fd}. In presence of a thermomagnetic medium, the soft contribution to the damping rate is found to be reduced than that of the thermal one. For small propagation angle, the reduction of the damping rate is more compared to that of thermal medium. For higher momentum the damping rate approaches the thermal value as the temperature becomes the dominant scale as compared to the
strength of the magnetic field considered.
\begin{figure}[htp]
	\centering
	\includegraphics[width=8cm]{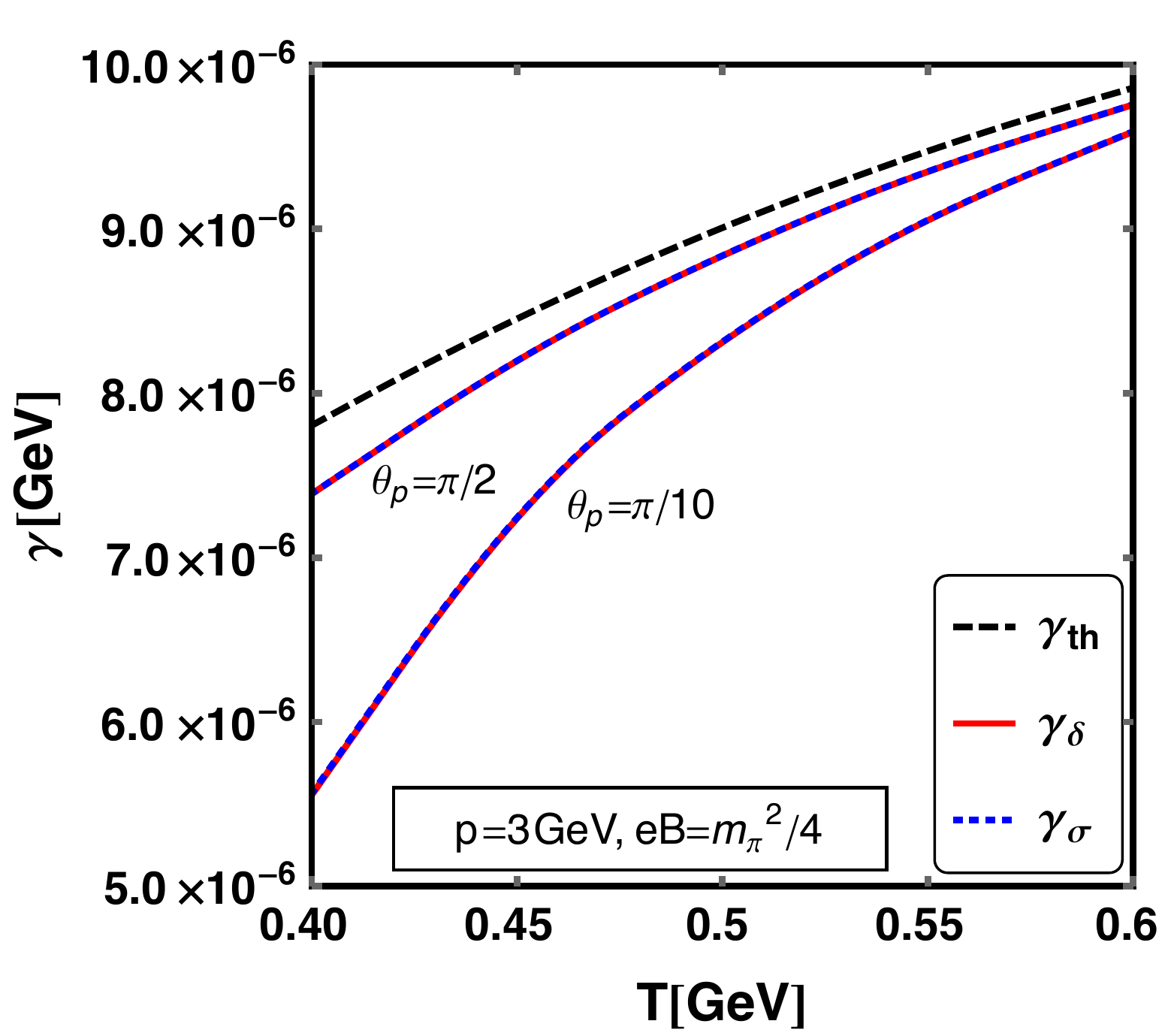}
	\caption{Plot of damping rate of the hard photon with temperature at $p=3$ GeV and $eB=m_\pi^2/4$ for two propagation angles $\pi/10$ and $\pi/2$.}
	\label{drvsT}
\end{figure}

Figure~\ref{drvsT} displays the variation of damping rate with temperature for a specific value of momentum and magnetic field for two propagation angles $\pi/10$ and $\pi/2$. It is found that the soft contribution to the damping rate increases with the increase in temperature both in thermal and thermomagnetic medium. For small propagation angle the damping rate is more reduced compared to that of large propagation angle. This observation is consistent with Fig.~\ref{drvsp}.
\begin{figure}[htp]
	\centering
	\includegraphics[width=8cm]{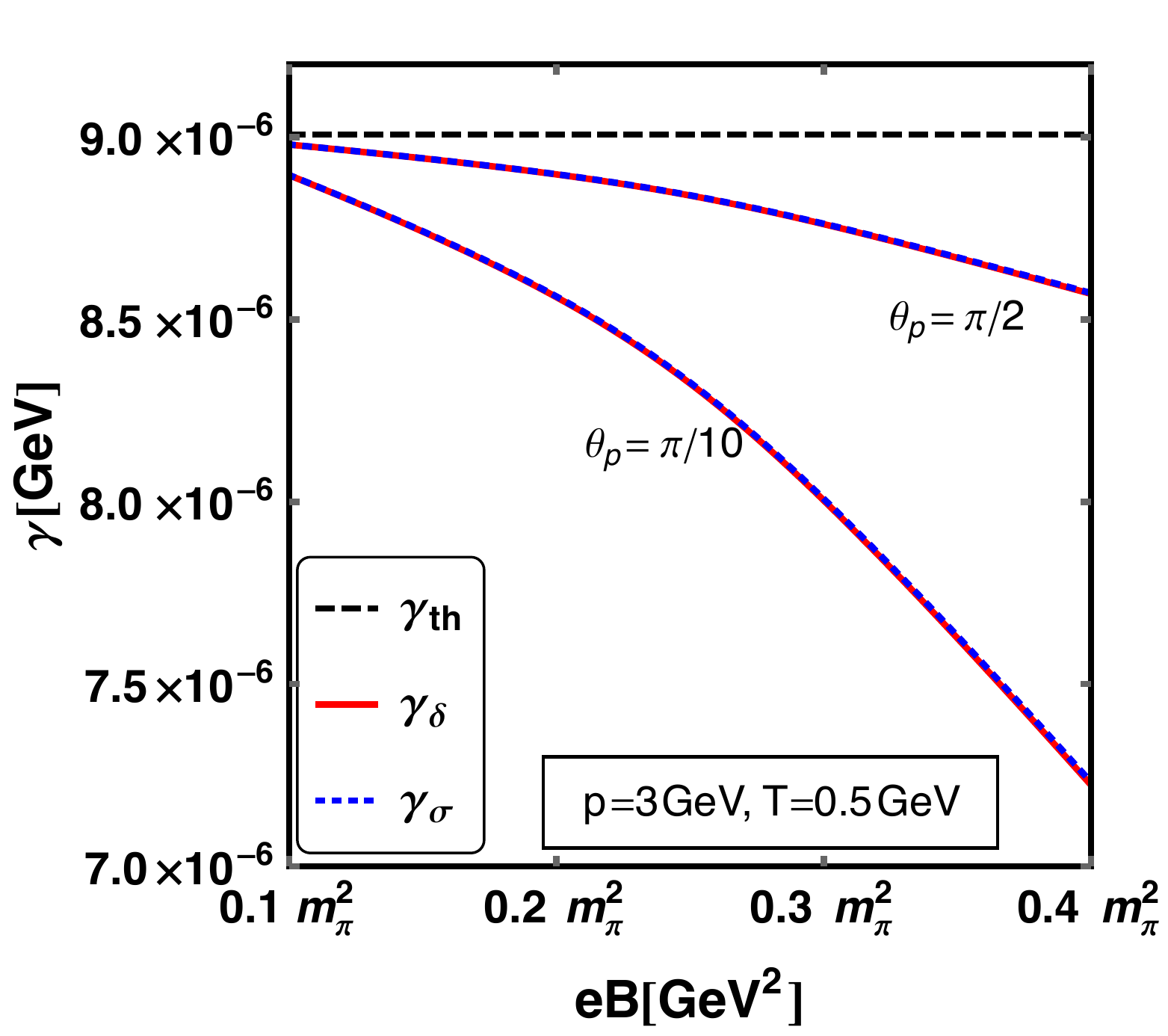}
	\caption{Plot of damping rate of the hard photon with the magnetic field strength at $T=0.5$ GeV and $p=3$ GeV for two propagation angles $\pi/10$ and $\pi/2$.}
	\label{drvseB}
\end{figure}

Figure~\ref{drvseB} shows the variation of the damping rate with the magnetic field strength for specific values of photon momentum and temperature for two propagation angles. The thermal damping rate ($\mathcal O[(eB)^0]$) is represented by the black dashed horizontal line. The thermomagnetic damping rate decreases with the increasing magnetic field. At smaller propagation angles the photons are less damped than that of higher propagation angles which are consistent with   Fig.~\ref{drvsp}.

\begin{figure}[htp]
	\centering
	\includegraphics[width=8cm]{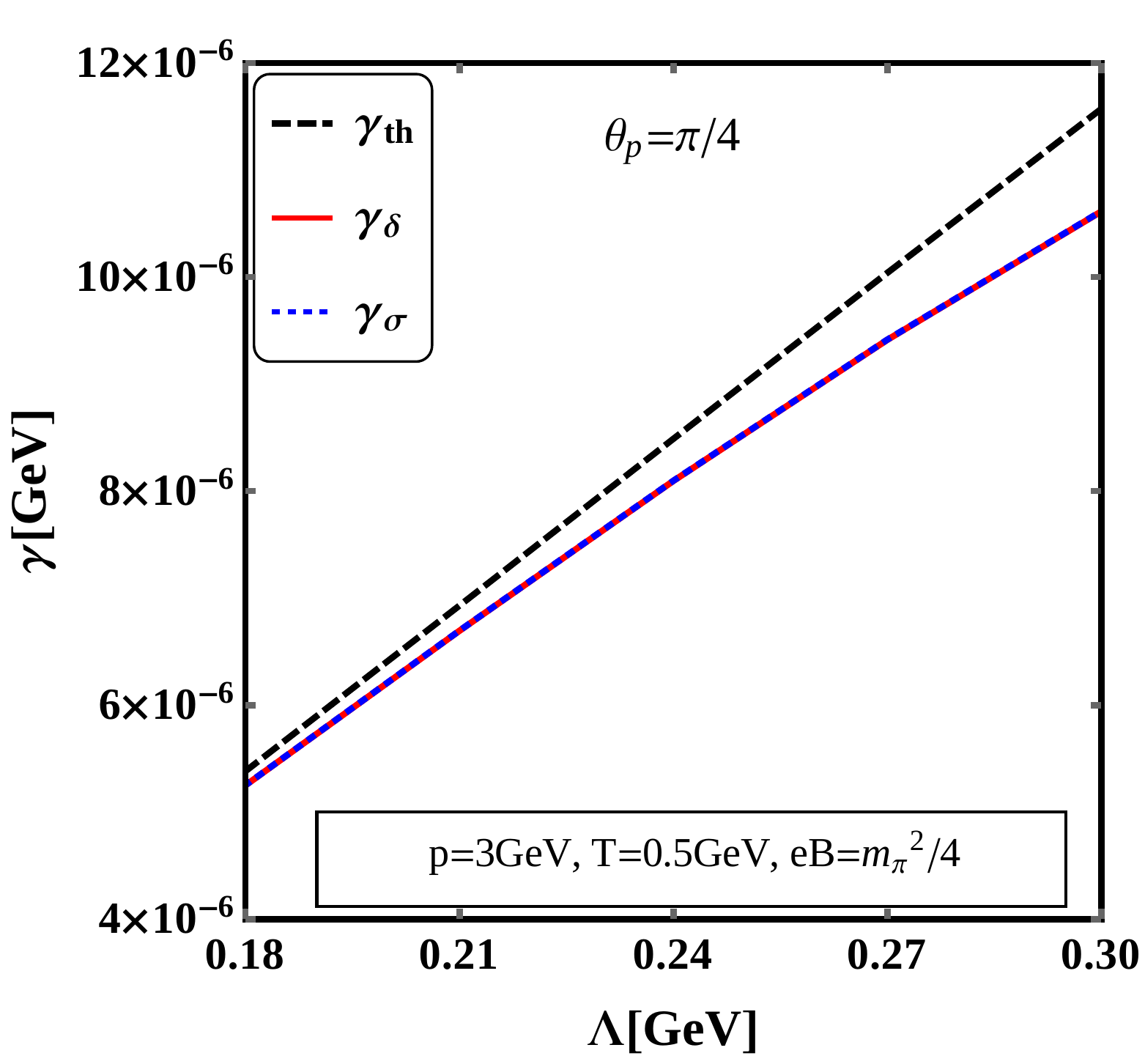}
	\caption{Plot of damping rate of photon with $\Lambda$ for $\theta_p=\pi/4$, $p=3$ GeV, $T=0.5$ GeV and $eB=m_\pi^2/4$.}
	\label{drvslambda}
\end{figure}
Fig.~\ref{drvslambda} shows the variation of the photon damping rate with the separation scale $\Lambda$ keeping the scale hierarchy $eT \ll \Lambda \ll T$. As the allowed phase space increases with the increase of $\Lambda$, the damping rate is also found to increase with it \footnote{Nevertheless, the damping rate is expected to be $\Lambda$ independent when hard contribution is added.}. The magnetic correction to the thermal damping rate is negative. So, the difference between the thermal and thermomagnetic damping rate increases with $\Lambda$.

\section{Conclusion}
\label{Conclusion}
We have calculated the soft contribution to the damping rate of a hard photon in a weakly magnetized QED medium where momentum of one of the fermion in the loop 
 is considered as soft. The two degenerate transverse modes of photon in thermal medium are damped in a similar fashion in presence of weak magnetic field as shown in Fig.~\ref{drvsang}. The difference between two transverse modes is very marginal due to weak field approximation.  The soft contribution to the damping rate in thermomagnetic medium  is reduced compared to that of thermal medium. When the magnetic field is switched off thermomagnetic damping modes reduce to its thermal value. The effect of magnetic field is found to be dominant at low temperature and low photon momentum.  
 
 The soft contribution to the hard photon damping rate is $\sim$ $10^{-6}$ GeV. Thus, a
photon of a few GeV energy traversing in the QED medium of temperature $\sim$$0.5$ GeV and background magnetic field $\sim$$0.005$ GeV$^2$
has a mean free path ($\lambda=\gamma^{-1}/2$) of a few \AA.
When the present calculation is extended to the case of relativistic heavy ion collisions, the mean free path of photon is found to be of a few hundred fm. This confirms that
the mean free path of photon is larger than the size of the fireball and photon can be treated as a direct probe.

The damping rate is found to be dependent on the separation scale $\Lambda$.  One needs to add the hard contribution with the soft contribution to cancel the $\Lambda$ dependence of the result.  The hard contribution to the photon damping rate comes from two-loop order with hard particles in the loop having momentum of the order of or higher than the temperature. This  itself is a huge calculation which is in progress.
\vspace{0.2in}

\noindent {\bf Acknowledgement:} 
RG is funded by University Grants Commission (UGC). BK and MGM were funded by Department of Atomic Energy (DAE), India via the 
project TPAES. RG and BK would like to thank Aritra Das for useful discussions.

\appendix
\section{Spectral representation of the propagators}
\label{app_a}
\bea
\lim_{\eps\to 0} \int_{-\infty}^{\infty}dx\,\frac{\eps}{x^2+\eps^2} \,\,f(x)&\approx&f(0)\int_{-\infty}^{\infty}dx\,\frac{\eps}{x^2+\eps^2}
\begin{cases}
	\rm{significant \ contribution\ comes\ from} \\ {\rm{integration\ ,where}} \,  x\simeq 0
\end{cases}
\nn\\
&=&f(0)\,\eps \int_{-\infty}^{\infty}dx\,\frac{1}{x^2+\eps^2} =\pi \, f(0) ,
\eea
where $f(x)$ is a test function.

\noindent
From the above equations we can write,
\bea
\lim_{\eps\to 0}\frac{\eps}{x^2+\eps^2}=\pi \delta(x),
\eea
\bea
\lim_{\eps\to 0} \text{Im}\, \frac{1}{x+i \eps}=\frac{1}{2i}\lim_{\eps\to 0}\bigg[\frac{1}{x+i\eps}-\frac{1}{x-i\eps}\bigg]=\frac{1}{2i}\lim_{\eps\to 0}\frac{-2 i \eps}{x^2+\eps^2}=-\pi \delta(x).
\label{Imx}
\eea

\bea
\lim_{\eps\to 0} \int_{-\infty}^{\infty}dx\,\frac{2 \eps x}{(x^2+\eps^2)^2}\, f(x) &=&\lim_{\eps\to 0} \int_{-\infty}^{\infty} dx\,\eps f(x)\frac{d}{dx}\bigg[-\frac{1}{(x^2+\eps^2)}\bigg] \nn\\
&=&\lim_{\eps\to 0}\int_{-\infty}^{\infty}dx\,f'(x)\frac{\eps}{x^2+\eps^2}
\nn\\
&=&\pi \,f'(0)=\pi \int dx\,f'(x)\delta(x)\nn\\
&=& -\pi \int dx\,f(x)\delta'(x)
\eea

\noindent
From the above equation we find,
\bea
\lim_{\eps\to 0}\frac{2\eps\,x}{(x^2+\eps^2)^2}=-\pi \delta'(x).
\label{Imx2}
\eea
\noindent
Now using Eq.~\eqref{Imx2} one can calculate,
\bea
\lim_{\eps\to 0}\text{Im}\,\frac{1}{(x+i\eps)^2}=\frac{1}{2i}\lim_{\eps\to 0}\bigg[\frac{1}{(x+i\eps)^2}-\frac{1}{(x-i \eps)^2}\bigg]=\frac{1}{2i}\lim_{\eps\to 0}\frac{-i4\eps x}{(x^2+\eps^2)^2}=\pi \delta'(x).
\eea

\noindent
Similarly,
\bea
\lim_{\eps\to 0}\int_{-\infty}^{\infty}dx\,\frac{\eps^3-3 x^2 \eps}{(x^2+\eps^2)^3}\,f(x)&=&\lim_{\eps\to 0}\int_{-\infty}^{\infty}dx\,\frac{\eps^3}{(x^2+\eps^2)^3}\,f(x)-\lim_{\eps\to 0}\int_{-\infty}^{\infty}dx\,\frac{3 x^2 \eps}{(x^2+\eps^2)^3}\,f(x)\nn\\
&=& I_1+I_2,
\eea
where
\bea
I_1&=&\lim_{\eps\to 0}\int_{-\infty}^{\infty}dx\,\frac{\eps^3\,f(x)}{(x^2+\eps^2)^3}\approx \lim_{\eps\to 0}\eps^3\, f(0)\int_{-\infty}^{\infty}dx\,\frac{1}{(x^2+\eps^2)^3}=\lim_{\eps\to 0}\,\eps^3 f(0)\frac{3}{8}\frac{1}{\eps^5}=\lim_{\eps\to 0}\frac{3}{8\eps^2}f(0),\nn\\
I_2&=&-3\lim_{\eps\to 0}\int_{-\infty}^{\infty}dx\,\frac{ x^2 \eps\,f(x)}{(x^2+\eps^2)^3}=-3\lim_{\eps\to 0}\,\eps \int dx\,f(x)\bigg[\frac{1}{8}\frac{d^2}{dx^2}\bigg(\frac{1}{x^2+\eps^2}\bigg)+\frac{1}{8}\frac{2}{(x^2+\eps^2)^2}\bigg]\nn\\
&=&- \lim_{\eps\to 0}\frac{3}{8\eps^2}f(0)-\frac{3}{8}\lim_{\eps\to 0}\eps\int dx\, f(x) \frac{d^2}{dx^2}\bigg(\frac{1}{x^2+\eps^2}\bigg).
\eea

So ,
\bea
I_1+I_2&=&-\frac{3}{8}\lim_{\eps\to 0}\eps\int dx\,f(x) \frac{d^2}{dx^2}\bigg(\frac{1}{x^2+\eps^2}\bigg)= \frac{3}{8}\lim_{\eps\to 0}\int dx\,\eps\,f'(x)\frac{d}{dx}\bigg(\frac{1}{x^2+\eps^2}\bigg) \nn\\
&=&- \frac{3}{8}\lim_{\eps\to 0}\int dx\,\eps\,f''(x)\frac{1}{x^2+\eps^2}=-\frac{3}{8}\pi \int dx\, f''(x)\delta(x)=-\frac{3}{8}\pi \int_{-\infty}^{\infty} dx\,f(x)\delta''(x).
\eea

We can conclude from the last few steps that,
\bea
\lim_{\eps\to 0}\frac{\eps^3-3 x^2\eps}{(x^2+\eps^2)^3}=-\frac{3}{8}\pi\delta''(x)\,\,.
\label{Imx3}
\eea
Using Eq.~\eqref{Imx3} we can find
\bea
\lim_{\eps\to 0}\text{Im}\frac{1}{(x+i\eps)^3}=\frac{1}{2i}\lim_{\eps\to 0}\bigg[\frac{1}{(x+i\eps)^2}-\frac{1}{(x-i\eps)^3}\bigg]=\lim_{\eps\to 0}\frac{\eps^3-6 \eps x^2}{(x^2+\eps^2)^3}=-\frac{3}{8}\pi \delta''(x)
\eea
Now ,
\bea
\lim_{\eps\to 0} \text{Im}\, \frac{1}{(x+i \eps)^4}=\frac{1}{2i}\lim_{\eps\to 0}\bigg[\frac{1}{(x+i\eps)^4}-\frac{1}{(x-i\eps)^4}\bigg]=\lim_{\eps\to 0}\frac{4x\eps^3-4x^3\eps}{(x^2+\eps^2)^4},
\eea
\bea
\lim_{\eps\to 0}&&\int_{-\infty}^{\infty}dx\,f(x)\frac{4x\eps^3-4x^3\eps}{(x^2+\eps^2)^4}=\lim_{\eps\to 0} 4\eps \int_{-\infty}^{\infty}dx\,f(x) \frac{1}{24}\frac{d^3}{dx^3}\bigg(\frac{1}{x^2+\eps^2}\bigg)\nn\\
&=&-\lim_{\eps\to 0} \frac{\eps}{6}\int_{-\infty}^{\infty}dx\, f'''(x)\frac{1}{x^2+\eps^2}=-\frac{\pi f'''(0)}{6}=\frac{\pi}{6}\int_{-\infty}^{\infty}dx\,f(x) \delta'''(x).
\eea
\bea
\mbox{Thus,\,\,\,}
\lim_{\eps\to 0} \text{Im}\, \frac{1}{(x+i \eps)^4}=\frac{\pi}{6}\delta'''(x).
\eea
\noindent
Now we write the spectral representations for the free propagators as
\bea
\rho_0^{(1)}(\om',q)&=& \frac{1}{2\pi}\lim_{\eps\to 0}\text{Im}\bigg[\frac{1}{\om'+i\eps + \omega_q}+\frac{1}{\om'+i\eps - \omega_q}\bigg]
=-\frac{1}{2}\bigg[\delta(\om'+\omega_q)+\delta(\om'-\omega_q)\bigg],\\
\rho_0^{(0)}(\om',q)&=& \frac{1}{2\pi \omega_q}\lim_{\eps\to 0}\text{Im}\bigg[\frac{1}{\om'+i\eps- \omega_q}-\frac{1}{\om'+i\eps + \omega_q}\bigg]
=\frac{1}{2\omega_q}\bigg[\delta(\om'+\omega_q)-\delta(\om'-\omega_q)\bigg],\\
\rho_1^{(1)}(\om',q)&=& \frac{1}{4 \pi \omega_q}\lim_{\eps\to 0}\text{Im}\bigg[\frac{1}{(\om'+i\eps - \omega_q)^2}-\frac{1}{(\om'+i\eps + \omega_q)^2}\bigg]\nn\\
&=&-\frac{1}{4\omega_q}\bigg[\delta'(\om'+\omega_q)-\delta'(\om'-\omega_q)\bigg],\\
\rho_1^{(0)}(\om',q)&=& \frac{1}{4\pi \omega_q^2}\lim_{\eps\to 0}\text{Im}\bigg[\frac{1}{(\om'+i\eps + \omega_q)^2}+\frac{1}{(\om'+i\eps - \omega_q)^2}-\frac{1}{\omega_q}\bigg(\frac{1}{\om'+i\eps - \omega_q}\nn\\
&-&\frac{1}{\om'+i\eps + \omega_q}\bigg)\bigg]\nn\\
&=&\frac{1}{4\omega_q^2}\bigg[\delta'(\om'+\omega_q)+\delta'(\om'-\omega_q)-\frac{1}{\omega_q}\bigg(\delta(\om'+\omega_q)-\delta(\om'-\omega_q)\bigg)\bigg],\\
\rho_2^{(1)}(\om',q)&=& \frac{1}{8 \pi \omega_q^2}\lim_{\eps\to 0}\text{Im}\bigg[ \frac{1}{(\om'+i\eps + \omega_q)^3} +\frac{1}{(\om'+i\eps - \omega_q)^3}+\frac{1}{2\omega_q}\bigg\{ \frac{1}{(\om'+i\eps + \omega_q)^2}\nn\\
&-&\frac{1}{(\om'+i\eps - \omega_q)^2}\bigg\}\bigg]\nn\\
&=&\frac{1}{8  \omega_q^2}\bigg[-\frac{3}{8}\bigg(\delta''(\om'+\omega_q)+\delta''(\om'-\omega_q) \bigg)+ \frac{1}{2\omega_q}\bigg( \delta'(\om'+\omega_q) -\delta'(\om'-\omega_q)\bigg) \bigg],\\
\rho_2^{(0)}(\om',q)&=&-\frac{1}{8 \pi \omega_q^3}\lim_{\eps\to 0}\text{Im}\bigg[\frac{1}{(\om'+i\eps + \omega_q)^3} -\frac{1}{(\om'+i\eps - \omega_q)^3} +\frac{1}{2\omega_q}\bigg\{\frac{1}{(\om'+i\eps + \omega_q)^2}\nn\\
&+&\frac{1}{(\om'+i\eps - \omega_q)^2} +\frac{3}{\omega_q}\bigg( \frac{1}{\om'+i\eps + \omega_q} -\frac{1}{\om'+i\eps - \omega_q} \bigg)\bigg\} \bigg]\nn\\
&=&-\frac{1}{8  \omega_q^3}\bigg[ -\frac{3}{8}\bigg(\delta''(\om'+\omega_q)-  \delta''(\om'-\omega_q)\bigg) +\frac{1}{2\omega_q}\bigg\{ \delta'(\om'+\omega_q)+\delta'(\om'-\omega_q)\nn\\
&-&\frac{3}{\omega_q}\bigg(\delta(\om'+\omega_q)-\delta(\om'-\omega_q)  \bigg)\bigg\}\bigg],\\
\rho_3^{(1)}(\om',q)&=&  -\frac{1}{16 \pi \omega_q^3}\lim_{\eps\to 0}\text{Im}\bigg[\frac{1}{(\om'+i\eps + \omega_q)^4} -\frac{1}{(\om'+i\eps - \omega_q)^4} +\frac{1}{\omega_q}\bigg\{\frac{1}{(\om'+i\eps + \omega_q)^3} \nn\\
&+&\frac{1}{(\om'+i\eps - \omega_q)^3} +\frac{1}{2\omega_q}\bigg( \frac{1}{(\om'+i\eps + \omega_q)^2} -\frac{1}{(\om'+i\eps - \omega_q)^2} \bigg)\bigg\} \bigg]\\
&=&-\frac{1}{16 \omega_q^3}\bigg[ \frac{1}{6}\bigg(\delta'''(\om'+\omega_q)-\delta'''(\om'-\omega_q)\bigg)+\frac{1}{\omega_q}\bigg\{ -\frac{3}{8}\bigg(\delta''(\om'+\omega_q)\nn\\
&+& \delta''(\om'-\omega_q)\bigg) +\frac{1}{2\omega_q} \bigg( \delta'(\om'+\omega_q)-\delta'(\om'-\omega_q)\bigg) \bigg\} \bigg],\\
\rho_3^{(0)}(\om',q)&=&  \frac{1}{16 \pi \omega_q^4}\lim_{\eps\to 0}\text{Im}\bigg[\frac{1}{(\om'+i\eps + \omega_q)^4} +\frac{1}{(\om'+i\eps - \omega_q)^4} +\frac{1}{2\omega_q}\bigg\{\frac{4}{(\om'+i\eps + \omega_q)^3} \nn\\
&-&\frac{4}{(\om'+i\eps - \omega_q)^3} +\frac{1}{2\omega_q}\bigg( \frac{10}{(\om'+i\eps + \omega_q)^2} +\frac{10}{(\om'+i\eps - \omega_q)^2} +\frac{1}{2\omega_q}\nn\\
&\times&\bigg( \frac{20}{\om'+i\eps+\omega_q}-\frac{20}{\om'+i\eps-\omega_q}\bigg)\bigg)\bigg\} \bigg]\\
&=&\frac{1}{16  \omega_q^4}\Bigg[\frac{1}{6}\bigg( \delta'''(\om'+\omega_q)+\delta'''(\om'-\omega_q)\bigg)+\frac{1}{2\omega_q}\bigg\{-\frac{3}{2}\bigg(\delta''(\om'+\omega_q)\nn\\
&-&\delta''(\om'-\omega_q) \bigg)+\frac{5}{\omega_q}\bigg( \delta'(\om'+\omega_q)+\delta'(\om'-\omega_q)-\frac{1}{\omega_q}\Big(\delta(\om'+\omega_q)\nn\\
&-&\delta(\om'-\omega_q) \Big)\bigg) \bigg\} \Bigg].
\eea
\noindent
The effective propagators are given as,
\bea
\frac{1}{D^2}=\frac{1}{D_+D_-}&=&   \sum_{i} \bigg(\frac{\partial D^2}{\partial \om}\bigg)^{-1}\bigg\vert_{\om=\om_i} \frac{1}{(\om-\om_i)} ,\\
\frac{1}{D^4}=\frac{1}{(D_+D_-)^2}&=&\sum_{i} \Bigg[\bigg(\frac{\partial D^2}{\partial \om}\bigg)^{-2}\bigg\vert_{\om=\om_i} \frac{1}{(\om-\om_i)^2}+ \frac{\partial}{\partial \om}\bigg\{\frac{(\om-\om_i)^2}{D^4} \bigg\}\bigg\vert_{\om=\om_i}\frac{1}{(\om-\om_i)}\Bigg]\nn\\
&=&\sum_{i} \Bigg[\bigg(\frac{\partial D^2}{\partial \om}\bigg)^{-2}\bigg\vert_{\om=\om_i} \frac{1}{(\om-\om_i)^2}-\frac{\partial^2 D^2 }{\partial \om^2}\bigg(\frac{1}{3!}\frac{\partial^3 D^{6}}{\partial \om^3}\bigg)^{-1}\Bigg\vert_{\om=\om_i}\frac{1}{(\om-\om_i)}\Bigg],\\
\frac{1}{D^6}=\frac{1}{(D_+D_-)^3}&=&\sum_{i} \Bigg[ \bigg(\frac{\partial D^2}{\partial \om}\bigg)^{-3}\bigg\vert_{\om=\om_i} \frac{1}{(\om-\om_i)^3}+ \frac{\partial}{\partial \om}\bigg\{\frac{(\om-\om_i)^3}{D^6} \bigg\}\bigg\vert_{\om=\om_i}\frac{1}{(\om-\om_i)^2} \nn\\
&+& \frac{1}{2}\frac{\partial^2}{\partial \om^2}\bigg\{\frac{(\om-\om_i)^3}{D^6} \bigg\}\bigg\vert_{\om=\om_i}\frac{1}{(\om-\om_i)}\Bigg]\nn\\
&=&\sum_{i} \Bigg[ \bigg(\frac{\partial D^2}{\partial \om}\bigg)^{-3}\bigg\vert_{\om=\om_i} \frac{1}{(\om-\om_i)^3}-\frac{3}{2}  \frac{\partial^2 D^2 }{\partial \om^2}\bigg(\frac{1}{4!}\frac{\partial^4 D^{8}}{\partial \om^4}\bigg)^{-1}\Bigg\vert_{\om=\om_i}\frac{1}{(\om-\om_i)^2} \nn\\
&-& \frac{3}{5}\Bigg(\frac{\partial^3 D^2}{\partial \om^3}\bigg\{ 6\bigg( \frac{\partial^4 D^8}{\partial \om^4} \bigg)^{-1}+\frac{7}{12}\bigg( \frac{\partial D^2}{\partial \om} \bigg)^{-4} \bigg\}+6\, \frac{\partial^2 D^2}{\partial \om^2} \frac{\partial}{\partial \om}\bigg( \frac{\partial^4 D^8}{\partial \om^4}\bigg)^{-1}\Bigg)\Bigg\vert_{\om=\om_i}\nn\\&\times&\frac{1}{(\om-\om_i)}\Bigg],
\eea
where $\om_i=\pm \om_{\pm}$ are the poles of $D_+$ and $D_-$.

\noindent
The spectral functions of the dressed propagators are given as
\bea
\rho_{D_{\pm}}&=& -\frac{(\om^2-k^2)}{2m_{\mathrm{th}}^2}\bigg[\delta(\om-\omega_\pm)+\delta(\om+\omega_\mp) \bigg]+\beta_{\pm}\Theta(k^2-\om^2),
\eea
where
\bea
\beta_{\pm}&=& \frac{-\frac{1}{2}(k\mp \om)m_{\mathrm{th}}^2}{\bigg[ k(\om\mp k)-m_{\mathrm{th}}^2\bigg\{ Q_0(\frac{\om}{k})\mp Q_1(\frac{\om}{k})\bigg\} \bigg]^2+\bigg[\frac{1}{2}(1\mp \frac{\om}{k}) m_{\mathrm{th}}^2 \pi \bigg]^2},
\eea
\noindent
where we use the Legendre function of  second kind
\bea
Q_0\bigg(\frac{\om}{k}\bigg)&=&\frac{1}{2}\ln\bigg|\frac{\om+k}{\om-k}\bigg|-i\frac{\pi}{2}\Theta(k^2-\om^2).
\eea
\bea
\rho_{4}(\om,k)&=&\frac{1}{\pi}\text{Im}\Big(\frac{1}{D^2} \Big)=- \sum_{i}  \bigg(\frac{\partial D^2}{\partial \om}\bigg)^{-1}\bigg\vert_{\om=\om_i} \delta(\om-\om_i)  + \beta_4 \Theta(k^2-\om^2)   \nn\\
&=&  \frac{\om^2-k^2}{4m_{\mathrm{th}}^2(k^2-\om^2+m_{\mathrm{th}}^2)}\bigg[(\om-k)\bigg(\delta(\om-\omega_+)+\delta(\om+\omega_-)\bigg)+(\om+k)\bigg(\delta(\om-\omega_-)\nn\\
&+&\delta(\om+\omega_+)\bigg) \bigg]+ \beta_4 \Theta(k^2-\om^2) ,\\
\rho_{5}(\om,k)&=&\frac{1}{\pi}\text{Im}\Big(\frac{1}{D^4} \Big)=\sum_{i} \Bigg[\bigg(\frac{\partial D^2}{\partial \om}\bigg)^{-2}\bigg\vert_{\om=\om_i} \delta'(\om-\om_i)+ \frac{\partial^2 D^2 }{\partial \om^2}\bigg(\frac{1}{3!}\frac{\partial^3 D^{6}}{\partial \om^3}\bigg)^{-1}\Bigg\vert_{\om=\om_i}\nn\\
&\times&\delta(\om-\om_i)\Bigg] + \beta_5 \Theta(k^2-\om^2),\\
\rho_{6}(\om,k)&=&\frac{1}{\pi}\text{Im}\Big(\frac{1}{D^6} \Big)=\sum_{i} \Bigg[-\frac{3}{8} \bigg(\frac{\partial D^2}{\partial \om}\bigg)^{-3}\bigg\vert_{\om=\om_i} \delta''(\om-\om_i)-\frac{3}{2} \frac{\partial^2 D^2 }{\partial \om^2}\bigg(\frac{1}{4!}\frac{\partial^4 D^{8}}{\partial \om^4}\bigg)^{-1}\Bigg\vert_{\om=\om_i}\nn\\
&\times&\delta'(\om-\om_i) + \frac{3}{5}\Bigg(\frac{\partial^3 D^2}{\partial \om^3}\bigg\{ 6\bigg( \frac{\partial^4 D^8}{\partial \om^4} \bigg)^{-1}+\frac{7}{12}\bigg( \frac{\partial D^2}{\partial \om} \bigg)^{-4} \bigg\}+6\, \frac{\partial^2 D^2}{\partial \om^2} \frac{\partial}{\partial \om}\bigg( \frac{\partial^4 D^8}{\partial \om^4}\bigg)^{-1}\Bigg)\Bigg\vert_{\om=\om_i}\nn\\
&\times&\delta(\om-\om_i)\Bigg] + \beta_6 \Theta(k^2-\om^2),\\
\rho_7 (\om,k)&=&\frac{1}{\pi}\text{Im}\bigg(\frac{b'}{D^2}\bigg)=-\,b' \sum_i  \bigg(\frac{\partial D^2}{\partial \om}\bigg\vert_{\om=\om_i} \bigg)^{-1}\delta(\om-\om_i)  + \beta_7 \Theta(k^2-\om^2), \\
\rho_8(\om,k)&=&\frac{1}{\pi}\text{Im}\bigg(\frac{c'}{D^2}\bigg)=-\,c' \sum_i  \bigg(\frac{\partial D^2}{\partial \om}\bigg\vert_{\om=\om_i} \bigg)^{-1}\delta(\om-\om_i)  + \beta_8 \Theta(k^2-\om^2) ,\\
\rho_9^{(0)}(\om,k)&=&\frac{1}{\pi}\text{Im}\bigg(\frac{h(1+a)}{D^4}\bigg)=h(1+a)\times \sum_i \Bigg[\bigg(\frac{\partial D^2}{\partial \om}\bigg\vert_{\om=\om_i} \bigg)^{-2}\delta'(\om-\om_i)\nn\\
&-& \frac{\partial}{\partial \om}\bigg\{\frac{(\om-\om_i)^2}{D^4} \bigg\}\bigg\vert_{\om=\om_i} \delta(\om-\om_i)\Bigg]+\beta_9 ^{(0)}\Theta(k^2-\om^2),\\
\rho_9^{(1)}(\om,k)&=&\frac{1}{\pi}\text{Im}\bigg(\frac{h(1+a)\om}{D^4}\bigg)=\om\,h(1+a)\times \sum_i \Bigg[\bigg(\frac{\partial D^2}{\partial \om}\bigg\vert_{\om=\om_i} \bigg)^{-2}\delta'(\om-\om_i)\nn\\
&-& \frac{\partial}{\partial \om}\bigg\{\frac{(\om-\om_i)^2}{D^4} \bigg\}\bigg\vert_{\om=\om_i} \delta(\om-\om_i)\Bigg]+\beta_9^{(1)} \Theta(k^2-\om^2),\\
\rho_{10}(\om,k)&=&\frac{1}{\pi}\text{Im}\bigg(\frac{h b}{D^4}\bigg)=h b \times \sum_i \Bigg[\bigg(\frac{\partial D^2}{\partial \om}\bigg\vert_{\om=\om_i} \bigg)^{-2}\delta'(\om-\om_i)- \frac{\partial}{\partial \om}\bigg\{\frac{(\om-\om_i)^2}{D^4} \bigg\}\bigg\vert_{\om=\om_i}\nn\\
&\times& \delta(\om-\om_i)\Bigg]+\beta_{10} \Theta(k^2-\om^2),\\
\rho_{11}(\om,k)&=&\frac{1}{\pi}\text{Im}\bigg[\frac{h b'}{D^4}\bigg]=h b'\times \sum_i \Bigg[\bigg(\frac{\partial D^2}{\partial \om}\bigg\vert_{\om=\om_i} \bigg)^{-2}\delta'(\om-\om_i)- \frac{\partial}{\partial \om}\bigg\{\frac{(\om-\om_i)^2}{D^4} \bigg\}\bigg\vert_{\om=\om_i}\nn\\
&\times& \delta(\om-\om_i)\Bigg]+\beta_{11} \Theta(k^2-\om^2),\\
\rho_{12}(\om,k)&=&\frac{1}{\pi}\text{Im}\bigg[\frac{h c'}{D^4}\bigg]=h c'\times \sum_i \Bigg[\bigg(\frac{\partial D^2}{\partial \om}\bigg\vert_{\om=\om_i} \bigg)^{-2}\delta'(\om-\om_i)- \frac{\partial}{\partial \om}\bigg\{\frac{(\om-\om_i)^2}{D^4} \bigg\}\bigg\vert_{\om=\om_i}\nn\\
&\times & \delta(\om-\om_i)\Bigg]+\beta_{12} \Theta(k^2-\om^2),\\
\rho_{13}(\om,k)&=&\frac{1}{\pi}\text{Im}\bigg[\frac{h'b}{D^4}\bigg]=h' b\times \sum_i \Bigg[\bigg(\frac{\partial D^2}{\partial \om}\bigg\vert_{\om=\om_i} \bigg)^{-2}\delta'(\om-\om_i)- \frac{\partial}{\partial \om}\bigg\{\frac{(\om-\om_i)^2}{D^4} \bigg\}\bigg\vert_{\om=\om_i}\nn\\
&\times& \delta(\om-\om_i)\Bigg]+\beta_{13} \Theta(k^2-\om^2),\\
\rho_{14}^{(0)}(\om,k)&=&\frac{1}{\pi}\text{Im}\bigg[\frac{h'(1+a)}{D^4}\bigg]=h' (1+a) \sum_i \Bigg[\bigg(\frac{\partial D^2}{\partial \om}\bigg\vert_{\om=\om_i} \bigg)^{-2}\delta'(\om-\om_i)\nn\\
&-& \frac{\partial}{\partial \om}\bigg\{\frac{(\om-\om_i)^2}{D^4} \bigg\}\bigg\vert_{\om=\om_i}\delta(\om-\om_i)\Bigg]+\beta_{14}^{(0)} \Theta(k^2-\om^2),\\
\rho_{14}^{(1)}(\om,k)&=&\frac{1}{\pi}\text{Im}\bigg[\frac{k_0\,h'(1+a)}{D^4}\bigg]=\om\,h' (1+a) \sum_i \Bigg[\bigg(\frac{\partial D^2}{\partial \om}\bigg\vert_{\om=\om_i} \bigg)^{-2}\delta'(\om-\om_i)\nn\\
&-& \frac{\partial}{\partial \om}\bigg\{\frac{(\om-\om_i)^2}{D^4} \bigg\}\bigg\vert_{\om=\om_i}\delta(\om-\om_i)\Bigg]+\beta_{14}^{(1)} \Theta(k^2-\om^2),\\
\rho_{15}^{(0)}(\om,k)&=&\frac{1}{\pi}\text{Im}\bigg[\frac{h^2 (1+a)}{D^6}\bigg]=h^2(1+a)\times \sum_i \Bigg[-\frac{3}{8} \bigg(\frac{\partial D^2}{\partial \om}\bigg\vert_{\om=\om_i} \bigg)^{-3}\delta''(\om-\om_i)\nn\\
&+& \frac{\partial}{\partial \om}\bigg\{\frac{(\om-\om_i)^3}{D^6} \bigg\}\bigg\vert_{\om=\om_i}\delta'(\om-\om_i) 
- \frac{\partial^2}{\partial \om^2}\bigg\{\frac{(\om-\om_i)^3}{D^6} \bigg\}\bigg\vert_{\om=\om_i}\delta(\om-\om_i)\Bigg]\nn\\
&+& \beta_{15}^{(0)} \Theta(k^2-\om^2),\\
\rho_{15}^{(1)}(\om,k)&=&\frac{1}{\pi}\text{Im}\bigg[\frac{h^2 (1+a)k_0}{D^6}\bigg]=\om\,h^2(1+a)\times \sum_i \Bigg[-\frac{3}{8} \bigg(\frac{\partial D^2}{\partial \om}\bigg\vert_{\om=\om_i} \bigg)^{-3}\delta''(\om-\om_i)\nn\\
&+& \frac{\partial}{\partial \om}\bigg\{\frac{(\om-\om_i)^3}{D^6} \bigg\}\bigg\vert_{\om=\om_i}\delta'(\om-\om_i) 
- \frac{\partial^2}{\partial \om^2}\bigg\{\frac{(\om-\om_i)^3}{D^6} \bigg\}\bigg\vert_{\om=\om_i}\delta(\om-\om_i)\Bigg]\nn\\
&+& \beta_{15}^{(1)} \Theta(k^2-\om^2),\\
\rho_{16}(\om,k)&=&\frac{1}{\pi}\text{Im}\bigg[\frac{h^2 b}{D^6}\bigg]=h^2b \times \sum_i \Bigg[-\frac{3}{8} \bigg(\frac{\partial D^2}{\partial \om}\bigg\vert_{\om=\om_i} \bigg)^{-3}\delta''(\om-\om_i)+ \frac{\partial}{\partial \om}\bigg\{\frac{(\om-\om_i)^3}{D^6} \bigg\}\bigg\vert_{\om=\om_i}\nn\\
&\times& \delta'(\om-\om_i)- \frac{\partial^2}{\partial \om^2}\bigg\{\frac{(\om-\om_i)^3}{D^6} \bigg\}\bigg\vert_{\om=\om_i}\delta(\om-\om_i)\Bigg] + \beta_{16} \Theta(k^2-\om^2),
\eea

where cut parts of the spectral functions are given as
\bea
\beta_4&=& \frac{1}{\pi}\text{Im}\bigg( \frac{1}{D^2}\bigg)=-\frac{1}{\pi}\frac{\text{Im}\,D^2 }{(\text{Re}\,D^2)^2+(\text{Im}\,D^2)^2},\\
\beta_5&=&\frac{1}{\pi}\text{Im}\bigg( \frac{1}{D^4}\bigg)=-\frac{1}{\pi}\frac{2\,\text{Re}\,D^2\, \text{Im}\,D^2 }{\Big[(\text{Re}\,D^2)^2+(\text{Im}\,D^2)^2\Big]^2},\\
\beta_6&=&\frac{1}{\pi}\text{Im}\bigg( \frac{1}{D^6}\bigg)=\frac{1}{\pi}\frac{\Big( \text{Im}\,D^2 \Big)^3-3\Big(\text{Re}\,D^2\Big)^2\,\text{Im}\,D^2 }{\Big[(\text{Re}\,D^2)^2+(\text{Im}\,D^2)^2\Big]^3},\\
\beta_7&=&\frac{1}{\pi}\text{Im}\bigg(\frac{b'}{D^2}\bigg)=\frac{1}{\pi}\frac{-\text{Re}\,b'\,\text{Im}\, D^2+\text{Im}\,b'\,\text{Re}\,D^2}{(\text{Re}\,D^2)^2+(\text{Im}\,D^2)^2},\\
\beta_8&=&\frac{1}{\pi}\text{Im}\bigg(\frac{c'}{D^2}\bigg)=\frac{1}{\pi}\frac{-\text{Re}\,c'\,\text{Im}\, D^2+\text{Im}\,c'\,\text{Re}\,D^2}{(\text{Re}\,D^2)^2+(\text{Im}\,D^2)^2},\\
\beta_9^{(0)}&=&\frac{1}{2}\text{Im}\bigg(\frac{h(1+a)}{D^4}\bigg)\nn\\
&=&\frac{1}{\pi}\frac{\text{Im}\,\big(h(1+a)\big)\bigg[(\text{Re}\,D^2)^2-(\text{Im}\,D^2)^2\bigg]-2\,\text{Re}\,D^2\, \text{Im}\,D^2\,\text{Re}\big(h(1+a)\big)}{\Big[(\text{Re}\,D^2)^2+(\text{Im}\,D^2)^2\Big]^2},\\
\beta_9^{(1)}&=&\frac{1}{2}\text{Im}\bigg(\frac{h(1+a)k_0}{D^4}\bigg)\nn\\
&=&\frac{1}{\pi}\frac{\text{Im}\,\big(hk_0(1+a)\big)\bigg[(\text{Re}\,D^2)^2-(\text{Im}\,D^2)^2\bigg]-2\,\text{Re}\,D^2\, \text{Im}\,D^2\,\text{Re}\big(hk_0(1+a)\big)}{\Big[(\text{Re}\,D^2)^2+(\text{Im}\,D^2)^2\Big]^2},\\
\beta_{10}&=&\frac{1}{2}\text{Im}\bigg(\frac{h\,b}{D^4}\bigg)=\frac{1}{\pi}\frac{\text{Im}\,\big(h\,b\big)\bigg[(\text{Re}\,D^2)^2-(\text{Im}\,D^2)^2\bigg]-2\,\text{Re}\,D^2\, \text{Im}\,D^2\,\text{Re}\big(h\,b\big)}{\Big[(\text{Re}\,D^2)^2+(\text{Im}\,D^2)^2\Big]^2},\\
\beta_{11}&=&\frac{1}{2}\text{Im}\bigg(\frac{h\,b'}{D^4}\bigg)=\frac{1}{\pi}\frac{\text{Im}\,\big(h\,b'\big)\bigg[(\text{Re}\,D^2)^2-(\text{Im}\,D^2)^2\bigg]-2\,\text{Re}\,D^2\, \text{Im}\,D^2\,\text{Re}\big(h\,b'\big)}{\Big[(\text{Re}\,D^2)^2+(\text{Im}\,D^2)^2\Big]^2},\\
\beta_{12}&=&\frac{1}{2}\text{Im}\bigg(\frac{h\,c'}{D^4}\bigg)=\frac{1}{\pi}\frac{\text{Im}\,\big(h\,c'\big)\bigg[(\text{Re}\,D^2)^2-(\text{Im}\,D^2)^2\bigg]-2\,\text{Re}\,D^2\, \text{Im}\,D^2\,\text{Re}\big(h\,c'\big)}{\Big[(\text{Re}\,D^2)^2+(\text{Im}\,D^2)^2\Big]^2},\\
\beta_{13}&=&\frac{1}{2}\text{Im}\bigg(\frac{h'\,b}{D^4}\bigg)=\frac{1}{\pi}\frac{\text{Im}\,\big(h'\,b\big)\bigg[(\text{Re}\,D^2)^2-(\text{Im}\,D^2)^2\bigg]-2\,\text{Re}\,D^2\, \text{Im}\,D^2\,\text{Re}\big(h'\,b\big)}{\Big[(\text{Re}\,D^2)^2+(\text{Im}\,D^2)^2\Big]^2},\\
\beta_{14}^{(0)}&=&\frac{1}{2}\text{Im}\bigg(\frac{h'\,(1+a)}{D^4}\bigg)\nn\\
&=&\frac{1}{\pi}\frac{\text{Im}\,\big(h'\,(1+a)\big)\bigg[(\text{Re}\,D^2)^2-(\text{Im}\,D^2)^2\bigg]-2\,\text{Re}\,D^2\, \text{Im}\,D^2\,\text{Re}\big(h'\,(1+a)\big)}{\Big[(\text{Re}\,D^2)^2+(\text{Im}\,D^2)^2\Big]^2},\\
\beta_{14}^{(1)}&=&\frac{1}{2}\text{Im}\bigg(\frac{h'\,(1+a)k_0}{D^4}\bigg)\nn\\
&=&\frac{1}{\pi}\frac{\text{Im}\,\big(h'\,(1+a)k_0\big)\bigg[(\text{Re}\,D^2)^2-(\text{Im}\,D^2)^2\bigg]-2\,\text{Re}\,D^2\, \text{Im}\,D^2\,\text{Re}\big(h'\,(1+a)k_0\big)}{\Big[(\text{Re}\,D^2)^2+(\text{Im}\,D^2)^2\Big]^2},\nn\\\\
\beta_{15}^{(0)}&=&\frac{1}{\pi}\text{Im}\bigg(\frac{h^2(1+a)}{D^6}\bigg)\nn\\
&=&\frac{1}{\pi}\frac{\text{Im}\,(h^2\,(1+a))\bigg[(\text{Re}\,D^2)^3-3\text{Re}\,D^2(\text{Im}D^2)^2\bigg]+\text{Re}\,(h^2\,(1+a))\bigg[(\text{Im}\,D^2)^3-3\text{Im}\,D^2(\text{Re}\,D^2)^2\bigg]}{\Big[(\text{Re}\,D^2)^2+(\text{Im}\,D^2)^2\Big]^3}\nn\\\\
\beta_{15}^{(1)}&=&\frac{1}{\pi}\text{Im}\bigg(\frac{h^2(1+a)k_0}{D^6}\bigg)\nn\\
&=&\frac{1}{\pi}\frac{\text{Im}\,(h^2\,(1+a)k_0)\bigg[(\text{Re}\,D^2)^3-3\text{Re}\,D^2(\text{Im}D^2)^2\bigg]+\text{Re}\,(h^2\,(1+a)k_0)\bigg[(\text{Im}\,D^2)^3-3\text{Im}\,D^2(\text{Re}\,D^2)^2\bigg]}{\Big[(\text{Re}\,D^2)^2+(\text{Im}\,D^2)^2\Big]^3}\nn\\\\
\beta_{16}&=&\frac{1}{\pi}\text{Im}\bigg( \frac{h^2\,b}{D^6}\bigg)\nn\\
&=&\frac{1}{\pi}\frac{\text{Im}\,(h^2\,b)\bigg[(\text{Re}\,D^2)^3-3\text{Re}\,D^2(\text{Im}D^2)^2\bigg]+\text{Re}\,(h^2\,b)\bigg[(\text{Im}\,D^2)^3-3\text{Im}\,D^2(\text{Re}\,D^2)^2\bigg]}{\Big[(\text{Re}\,D^2)^2+(\text{Im}\,D^2)^2\Big]^3}\nn\\
\eea
where
\bea
\text{Im}(D^2)&=&-\frac{\pi m_{\mathrm{th}}^4}{ k^2}\bigg[\frac{\om}{k}+\Big(1-\frac{\om^2}{k^2}\Big) Q_0\left(\frac{\om}{k}\right)\bigg],\\
\text{Re}(D^2)&=&\om^2-k^2-2m_{\mathrm{th}}^2+\frac{m_{\mathrm{th}}^4}{k^2}\bigg(\frac{2\om}{k}Q_0\bigg( \frac{\om}{k} \bigg)-1  \bigg)+\frac{m_{\mathrm{th}}^4}{k^2}\Big(1-\frac{\om^2}{k^2} \Big)\bigg( Q_0^2\bigg( \frac{\om}{k} \bigg)-\frac{\pi^2}{4} \bigg),\\
\text{Im}(b')&=&-4e^2 M^2\frac{\pi\, k_3\, \om}{2k^3},\\
\text{Re}(b')&=&4 e^2  M^2\frac{k_3}{k^2}\bigg[\frac{\om}{k}Q_0\bigg( \frac{\om}{k} \bigg)-1\bigg],\\
\text{Im}(c')&=&-4e^2 M^2\frac{\pi}{2k},\\
\text{Re}(c')&=&4e^2 M^2\frac{1}{k}Q_0\bigg( \frac{\om}{k} \bigg),\\
  \text{Im}\Big(h(1+a)\Big)&=&\frac{4\pi e^2 M^2   k_3}{2 k^7} \bigg(-2 k^6-2 k^4 \left(k_0^2+3 \,m_{\mathrm{th}}^2\right)+12 k^3 k_0 \,m_{\mathrm{th}}^2 Q_0-4 k^2 \,m_{\mathrm{th}}^2 \left(k_0^2+\,m_{\mathrm{th}}^2\right)\nn\\
  &+&4 k k_0 \,m_{\mathrm{th}}^2
   Q_0 \left(k_0^2+4 \,m_{\mathrm{th}}^2\right)+k_0^2 \,m_{\mathrm{th}}^4 \left(\pi ^2-12 Q_0^2\right)\bigg),\nn\\
\text{Re}(h(1+a))&=&\frac{4e^2M^2 k_3}{2 k^7} \bigg(4 k^6 Q_0-4 k^5 k_0+4 k^4 Q_0 \left(k_0^2+3 \,m_{\mathrm{th}}^2\right)+k^3 k_0 \,m_{\mathrm{th}}^2 \left(-12 Q_0^2+3 \pi ^2-4\right)\nn\\
&+&8 k^2 m_{\mathrm{th}}^2
   Q_0 \left(k_0^2+\,m_{\mathrm{th}}^2\right)+k k_0 \,m_{\mathrm{th}}^2 \left(\pi ^2-4 Q_0^2\right) \left(k_0^2+4 \,m_{\mathrm{th}}^2\right)+2 k_0^2 \,m_{\mathrm{th}}^4 Q_0 \nn\\
   &\times&\left(4 Q_0^2-3 \pi
   ^2\right)\bigg),\\
\text{Im}(h\,b)&=&\frac{  4\pi e^2M^2 k_3 \,m_{\mathrm{th}}^2 }{2 k^7}\bigg(4 k^2 k_0 \left(k_0^2+\,m_{\mathrm{th}}^2\right)+4 k Q_0 \left(k^4+2 \,m_{\mathrm{th}}^2 \left(k^2-2 k_0^2\right)-k_0^4\right)\nn\\
&+&12 k_0
   \,m_{\mathrm{th}}^2 Q_0^2 (k_0-k) (k+k_0)+\pi ^2 k_0 \,m_{\mathrm{th}}^2 (k-k_0) (k+k_0)\bigg),\\
\text{Re}(h\,b)&=&\frac{4e^2M^2 k_3 \,m_{\mathrm{th}}^2 }{2 k^7}\bigg(k^5 \left(\pi ^2-4 Q_0^2\right)+2 k^3 \left(2 k_0^2+\,m_{\mathrm{th}}^2 \left(\pi ^2-4 Q_0^2\right)\right)-2 k^2 k_0 Q_0\nn\\
&\times&\Big(4
   k_0^2+\,m_{\mathrm{th}}^2 \left(-4 Q_0^2+3 \pi ^2+4\right)\Big)-k k_0^2 \left(\pi ^2-4 Q_0^2\right) \left(k_0^2+4 \,m_{\mathrm{th}}^2\right)+2 k_0^3 \,m_{\mathrm{th}}^2 Q_0 \nn\\
   &\times&\left(3 \pi ^2-4
   Q_0^2\right)\bigg),\\
\text{Im}(h\,b')&=&\frac{16 \pi e^4M^4 k_3^2}{2 k^7} \bigg(2 k^4-4 k^3 k_0 Q_0+4 k^2 \left(k_0^2+\,m_{\mathrm{th}}^2\right)-4 k k_0 Q_0 \left(k_0^2+4 \,m_{\mathrm{th}}^2\right)\nn\\
&-&k_0^2 \,m_{\mathrm{th}}^2
   \left(\pi ^2-12 Q_0^2\right)\bigg),\\
\text{Re}(h\,b')&=&-\frac{16e^4M^4 k_3^2 }{2 k^7}\bigg(4 k^4 Q_0+k^3 k_0 \left(\pi ^2-4 \left(Q_0^2+1\right)\right)+8 k^2 Q_0 \left(k_0^2+\,m_{\mathrm{th}}^2\right)\nn\\
&+&k k_0 \left(\pi ^2-4
   Q_0^2\right) \left(k_0^2+4 \,m_{\mathrm{th}}^2\right)+2 k_0^2 \,m_{\mathrm{th}}^2 Q_0 \left(4 Q_0^2-3 \pi ^2\right)\bigg),\\
\text{Im}(h\,c')&=&-\frac{16 \pi e^4M^4 k_3}{2 k^5} \left(4 k^3 Q_0-2 k^2 k_0+4 k Q_0 \left(k_0^2+2 \,m_{\mathrm{th}}^2\right)+k_0 \,m_{\mathrm{th}}^2 \left(\pi ^2-12 Q_0^2\right)\right),\\
\text{Re}(h\,c')&=&-\frac{16e^4M^4 k_3}{2 k^5} \bigg(k^3 \left(\pi ^2-4 Q_0^2\right)+4 k^2 k_0 Q_0+k \left(\pi ^2-4 Q_0^2\right) \left(k_0^2+2 \,m_{\mathrm{th}}^2\right)\nn\\
&+&2 k_0 \,m_{\mathrm{th}}^2 Q_0
   \left(4 Q_0^2-3 \pi ^2\right)\bigg),\\
\text{Im}(h'b)&=&\frac{8 \pi  e^4 M^4 m_{\mathrm{th}}^2 }{k^9}\bigg(k^6 \left(\frac{\pi ^2}{4}-3 Q_0^2\right)-2 k^5 \om Q_0+k^4 \Big(k_3^2-\om^2
   \left(\frac{\pi ^2}{4}-3 Q_0^2\right)\Big)\nn\\
   &-&4 k^3 \om k_3^2 Q_0-k^2 \om^2 k_3^2 \left(-3 Q_0^2+\frac{\pi
   ^2}{4}+3\right)+6 k \om^3 k_3^2 Q_0+\om^4 k_3^2 \left(\frac{\pi ^2}{4}-3 Q_0^2\right)\bigg),\\
\text{Re}(h'b)&=&\frac{16  e^4 M^4 m_{\mathrm{th}}^2}{k^9} \Bigg(k^6 \left(Q_0^3-\frac{3 \pi ^2}{4} Q_0\right)+k^5 \om \left(Q_0^2-\frac{\pi ^2}{4}\right)-k^4 Q_0
   \Big(\om^2 \Big(Q_0^2-\frac{3 \pi ^2}{4}\Big)+k_3^2\Big)\nn\\
   &+&k^3 \om k_3^2 \left(2 Q_0^2-\frac{ \pi ^2}{2}-1\right)+k^2
   \om^2 k_3^2 Q_0 \left(-Q_0^2+\frac{3 \pi ^2}{4}+3\right)+3 k \om^3 k_3^2 \Big(\frac{\pi^2}{4} -Q_0^2\Big)\nn\\
   &+&\om^4 k_3^2 Q_0 \left(Q_0^2-\frac{3 \pi ^2}{4}\right)\Bigg),\\
\text{Im}(h'\,(1+a))&=&\frac{8 \pi  e^4 M^4 }{k^9}\bigg(2 k^7 Q_0+2 k^5 m_{\mathrm{th}}^2 Q_0+k^4 \om \Big(2 k_3^2+m_{\mathrm{th}}^2 \Big(\frac{\pi ^2}{4}-3
   Q_0^2\Big)\Big)-2 k^3 \om^2 k_3^2 Q_0\nn\\
   &+&3 k^2 \om k_3^2 m_{\mathrm{th}}^2-6 k \om^2 k_3^2
  m_{\mathrm{th}}^2 Q_0-\om^3 k_3^2 m_{\mathrm{th}}^2 \Big(\frac{\pi ^2}{4}-3 Q_0^2\Big)\bigg),\\
\text{Re}(h'\,(1+a))&=&\frac{16  e^4 M^4}{k^9} \bigg(k^7 \Big(\frac{\pi^2}{4} -Q_0^2\Big) +k^5 \Big(k_3^2+m_{\mathrm{th}}^2 \Big(\frac{\pi^2}{4} -Q_0^2\Big) \Big)+k^4
   \om Q_0 \nn\\
   &\times&\Big(m_{\mathrm{th}}^2 \Big(Q_0^2-\frac{3 \pi ^2}{4}\Big)-2 k_3^2\Big)+k^3 k_3^2 \left(\om^2
   \Big(Q_0^2-\frac{\pi ^2}{4}\Big)+m_{\mathrm{th}}^2\right)-3 k^2 \om k_3^2 m_{\mathrm{th}}^2 Q_0\nn\\
   &+&3 k \om^2 k_3^2 m_{\mathrm{th}}^2
   \left(Q_0^2-\frac{\pi ^2}{4}\right)-\om^3 k_3^2 m_{\mathrm{th}}^2 Q_0\Big(Q_0^2-\frac{3 \pi ^2}{4}\Big)\bigg),\\
\text{Im}\Big(h^2(1+a)\Big)&=&\frac{16 \pi e^4M^4 k_3^2 }{2 k^{11}}\Bigg(-8 k^9 Q_0+8 k^8 k_0-8 k^7 Q_0 \left(2 k_0^2+5 \,m_{\mathrm{th}}^2\right)+k^6 k_0 \bigg(8 k_0^2\nn\\
&+&\,m_{\mathrm{th}}^2 \left(60 Q_0^2-5 \pi
   ^2+24\right)\bigg)-8 k^5 Q_0 \left(k_0^4+12 k_0^2 \,m_{\mathrm{th}}^2+8 \,m_{\mathrm{th}}^4\right)+2 k^4 k_0 \,m_{\mathrm{th}}^2 \nn\\
   &\times&\bigg(k_0^2 \left(36 Q_0^2-3 \pi ^2+6\right)-8 \,m_{\mathrm{th}}^2
   \left(-12 Q_0^2+\pi ^2-1\right)\bigg)-8 k^3 \,m_{\mathrm{th}}^2 Q_0 \bigg(3 k_0^4\nn\\
   &-&4 k_0^2 \,m_{\mathrm{th}}^2 \left(-4 Q_0^2+\pi ^2-3\right)+4 \,m_{\mathrm{th}}^4\bigg)-k^2 k_0 \,m_{\mathrm{th}}^2
   \left(\pi ^2-12 Q_0^2\right) \Big(k_0^4+12 k_0^2 \,m_{\mathrm{th}}^2\nn\\
   &+&12 \,m_{\mathrm{th}}^4\Big)+16 k k_0^2 \,m_{\mathrm{th}}^4 Q_0 \left(\pi ^2-4 Q_0^2\right) \left(k_0^2+3
   \,m_{\mathrm{th}}^2\right)\nn\\
   &+&k_0^3 \,m_{\mathrm{th}}^6 \left(80 Q_0^4-40 \pi ^2 Q_0^2+\pi ^4\right)\Bigg),\\
\text{Re}\Big(h^2(1+a)\Big)&=&-\frac{16e^4M^4 k_3^2}{k^{11}} \Bigg(k^9 \left(\pi ^2-4 Q_0^2\right)+8 k^8 k_0 Q_0+k^7 \Big(2 k_0^2 \left(-4 Q_0^2+\pi ^2-2\right)\nn\\
&+&5 \,m_{\mathrm{th}}^2 \left(\pi ^2-4
   Q_0^2\right)\Big)+k^6 k_0 Q_0 \Big(8 k_0^2+\,m_{\mathrm{th}}^2 \left(20 Q_0^2-15 \pi ^2+24\right)\Big)\nn\\
   &+&k^5 \Big(k_0^4 \left(\pi ^2-4 Q_0^2\right)+4 k_0^2
   \,m_{\mathrm{th}}^2 \big(-12 Q_0^2+3 \pi ^2-1\big)+8 \,m_{\mathrm{th}}^4 \left(\pi ^2-4 Q_0^2\right)\Big)\nn\\
   &+&2 k^4 k_0 \,m_{\mathrm{th}}^2 Q_0 \bigg(k_0^2 \Big(12 Q_0^2-9 \pi ^2+6\Big)+8
   \,m_{\mathrm{th}}^2 \left(4 Q_0^2-3 \pi ^2+1\right)\bigg)\nn\\
   &+&k^3 \,m_{\mathrm{th}}^2 \bigg(3 k_0^4 \left(\pi ^2-4 Q_0^2\right)-2 k_0^2 \,m_{\mathrm{th}}^2 \Big(8 \left(2 Q_0^2+3\right) Q_0^2-6 \pi
   ^2 \left(4 Q_0^2+1\right)+\pi ^4\Big)\nn\\
   &+&4 \,m_{\mathrm{th}}^4\left(\pi ^2-4 Q_0^2\right)\bigg)+k^2 k_0 \,m_{\mathrm{th}}^2 Q_0 \left(4 Q_0^2-3 \pi ^2\right) \left(k_0^4+12 k_0^2
   \,m_{\mathrm{th}}^2+12 \,m_{\mathrm{th}}^4\right)\nn\\
   &-&k k_0^2 \,m_{\mathrm{th}}^4\left(16 Q_0^4-24 \pi ^2 Q_0^2+\pi ^4\right) \left(k_0^2+3 \,m_{\mathrm{th}}^2\right)\nn\\
   &+&k_0^3 \,m_{\mathrm{th}}^6 Q_0 \left(16
   Q_0^4-40 \pi ^2 Q_0^2+5 \pi ^4\right)\Bigg),\\
\text{Im}(h^2b)&=&\frac{16 \pi e^4M^4 k_3^2 \,m_{\mathrm{th}}^2 }{2 k^{11}}\Bigg(k^8 \left(-\left(\pi ^2-12 Q_0^2\right)\right)-8 k^7 k_0 Q_0-k^6 \Big(k_0^2 \left(-12 Q_0^2+\pi ^2+4\right)\nn\\
&+&4 \,m_{\mathrm{th}}^2
   \left(\pi ^2-12 Q_0^2\right)\Big)+16 k^5 k_0 Q_0 \left(k_0^2+\,m_{\mathrm{th}}^2 \left(\pi ^2-4 Q_0^2\right)\right)\nn\\
   &+&k^4 \bigg(k_0^4 \Big(\pi ^2-12
   \left(Q_0^2+1\right)\Big)-16 k_0^2 \,m_{\mathrm{th}}^2-4 \,m_{\mathrm{th}}^4 \left(\pi ^2-12 Q_0^2\right)\bigg)\nn\\
   &+&8 k^3 k_0 Q_0 \left(3 k_0^4+12 k_0^2 \,m_{\mathrm{th}}^2+4 \,m_{\mathrm{th}}^4
   \left(-4 Q_0^2+\pi ^2+1\right)\right)\nn\\
   &+&k^2 k_0^2 \bigg(k_0^4 \left(\pi ^2-12 Q_0^2\right)+12 k_0^2 \,m_{\mathrm{th}}^2 \left(\pi ^2-12 Q_0^2\right)+\,m_{\mathrm{th}}^4 \Big(80 Q_0^4-8
   \left(18+5 \pi ^2\right) Q_0^2\nn\\
   &+&\pi ^2 \big(12+\pi ^2\big)\Big)\bigg)-16 k k_0^3 \,m_{\mathrm{th}}^2 Q_0 \left(\pi ^2-4 Q_0^2\right) \left(k_0^2+3 \,m_{\mathrm{th}}^2\right)\nn\\
   &-&k_0^4
   \,m_{\mathrm{th}}^4 \left(80 Q_0^4-40 \pi ^2 Q_0^2+\pi ^4\right)\Bigg),\\   
\text{Re}(h^2 b)&=&\frac{16e^4M^4 k_3^2 \,m_{\mathrm{th}}^2 }{k^{11}}\Bigg(k^8 \left(3 \pi ^2 Q_0-4 Q_0^3\right)-k^7 k_0 \left(\pi ^2-4 Q_0^2\right)+k^6 Q_0 \Big(k_0^2 \left(-4 Q_0^2+3 \pi
   ^2+4\right)\nn\\
   &+&4 \,m_{\mathrm{th}}^2 \left(3 \pi ^2-4 Q_0^2\right)\Big)+k^5 k_0 \left(2 k_0^2 \left(-4 Q_0^2+\pi ^2-2\right)+\,m_{\mathrm{th}}^2 \left(16 Q_0^4-24 \pi ^2 Q_0^2+\pi
   ^4\right)\right)\nn\\
   &+&k^4 Q_0 \bigg(k_0^4 \left(4 Q_0^2-3 \pi ^2+12\right)+16 k_0^2 \,m_{\mathrm{th}}^2+4 \,m_{\mathrm{th}}^4 \left(3 \pi ^2-4 Q_0^2\right)\bigg)+k^3 k_0 \bigg(3 k_0^4
   \left(\pi ^2-4 Q_0^2\right)\nn\\
   &+&12 k_0^2 \,m_{\mathrm{th}}^2 \left(\pi ^2-4 Q_0^2\right)+2 \,m_{\mathrm{th}}^4 \left(16 Q_0^4-8 \left(1+3 \pi ^2\right) Q_0^2+\pi ^2 \left(2+\pi
   ^2\right)\right)\bigg)+k^2 k_0^2 Q_0\nn\\
   &\times&\bigg(4 Q_0^2 \left(k_0^4+12 k_0^2 \,m_{\mathrm{th}}^2+2 \left(6+5 \pi ^2\right) \,m_{\mathrm{th}}^4\right)-\pi ^2 \left(3 k_0^4+36 k_0^2
   \,m_{\mathrm{th}}^2+\left(36+5 \pi ^2\right) \,m_{\mathrm{th}}^4\right)\nn\\
   &-&16 \,m_{\mathrm{th}}^4 Q_0^4\bigg)-k k_0^3 \,m_{\mathrm{th}}^2 \big(16 Q_0^4-24 \pi ^2 Q_0^2+\pi ^4\big) \left(k_0^2+3
   \,m_{\mathrm{th}}^2\right)+k_0^4 \,m_{\mathrm{th}}^4 Q_0 \big(16 Q_0^4\nn\\
   &-&40 \pi ^2 Q_0^2+5 \pi ^4\big)\Bigg),
\eea
where $M^2$ is defined in Eq.~\eqref{mag_mass}.

\end{document}